\documentclass{article}
\usepackage[utf8]{inputenc}
\usepackage[english]{babel}
\usepackage{microtype}
\usepackage{ragged2e} 
\usepackage{graphicx} 
\usepackage{subcaption}
\usepackage{float}
\usepackage{bm}
\usepackage{mathtools}
\usepackage{amsmath,amsfonts,amssymb,amsthm}
\usepackage{fullpage}
\usepackage{pgfplots} 
\usepackage{multirow}
\usepackage[left = 1 in, right = 1 in, top = 1 in, bottom = 1 in]{geometry}

\newcommand{\indep}{\perp \!\!\! \perp}

\theoremstyle{plain}
\newtheorem{assumption}{Assumption}

\theoremstyle{plain}
\newtheorem{theorem}{Theorem}

\theoremstyle{plain}

\theoremstyle{plain}

\usepackage[nottoc,notlof,notlot]{tocbibind} 
\usepackage[titles]{tocloft} 

\usepackage[
backend=biber,
style=authoryear,
sorting=nyt,
maxbibnames=10, 
maxcitenames=3,  
]{biblatex}
\usepackage[hidelinks,colorlinks=true,linkcolor=blue,citecolor=blue]{hyperref}

\usepackage{setspace}
\usepackage[fontsize=12pt]{fontsize}

\title{\sc A Structural Approach to Growth-at-Risk\thanks{Acknowledgments: I am indebted to Christian Brownlees and Andrea Caggese for their support and guidance. I would also like to thank Geert Mesters, Barbara Rossi, Vladislav Morozov, Valeria Gargiulo and the participants of the 48th Symposium of the Spanish Economic Association and the Euro Area Business Cycle Network: Advances in Local Projections and Empirical Methods for Central Banking conference for comments and discussion. All remaining errors are my own.}}
\author{Robert Wojciechowski\thanks{Universitat Pompeu Fabra, robert.wojciechowski@upf.edu}}
\date{\today}

\begin{document}
\maketitle

\begin{abstract}

We identify the structural impulse responses of quantiles of the outcome variable to a shock. Our estimation strategy explicitly distinguishes treatment from control variables, allowing us to model responses of unconditional quantiles while using controls for identification. Disentangling the effect of adding control variables on identification versus interpretation brings our structural quantile impulse responses conceptually closer to structural mean impulse responses. Applying our methodology to study the impact of financial shocks on lower quantiles of output growth confirms that financial shocks have an outsized effect on growth-at-risk, but the magnitude of our estimates is more extreme than in previous studies.

\smallskip

{\setlength{\parindent}{0cm}
}

\end{abstract}

\bigskip

\pagebreak

\section{Introduction}

A large empirical literature that studies the relationship between financial conditions and output growth has documented that tightening financial conditions are associated with periods of growth vulnerability. During these periods, the distribution of output growth is left-skewed making recessions more likely \parencite{Adrian2019}.  In light of these findings, modeling how financial conditions affect the left-tail of the distribution of growth -- rather than just its central tendency -- has become increasingly important for academics and policymakers alike. A theoretical macrofinance literature rationalizes the heterogenous effect of financial conditions on upside versus downside risk with models of bank-runs and intermediation crises. Some of the key features of these models are occasionally binding constraints on financial intermediation, the interaction of shocks with bank balance sheet conditions and amplification mechanisms \parencite{He2019, Gertler2019, Brunnermeier2014}.

A commonly used measure for downside risk is growth-at-risk (GaR) -- defined as the $0.05$th quantile of the future output growth distribution. It is a pessimistic growth scenario, which materializes five percent of the time. GaR literature is primarily focused on forecasting \parencite{Adrian2019, Plagborg2020, Brownlees2021}, and policymakers use declining GaR forecasts as an early warning signal of a coming recession. There are also early attempts at studying the structural drivers of GaR \parencite{Adrian2020, Adrian2023} to guide macro-prudential policy and to evaluate the mechanisms put forth in the theoretical literature. 

Quantile regression of \cite{Koenker1978} is a natural methodology to study the determinants of GaR as it estimates quantile specific coefficients. For instance, we can study how GaR responds to financial conditions by estimating a quantile regression (for the $0.05$th quantile) of output growth on financial conditions. A drawback of quantile regression is that it estimates the slope coefficients of the \textit{conditional} quantiles of the outcome. Therefore, adding controls into the regression changes the conditioning set and thus also alters the interpretation of the estimated quantile specific coefficients. Note that this is not simply a case of omitted variable bias, as even the addition of covariates that are statistically independent can change the coefficients. This problem is not encountered in conditional expectations models, where the Fisch-Waugh-Lovell theorem applies, and the inclusion of additional uncorrelated covariates only affects the standard errors.

Adding macroeconomic controls to a quantile regression of output growth on financial conditions will alter the interpretation of the coefficient on financial conditions from ``the effect of financial conditions when growth is low" to ``the effect of a shock to financial conditions when growth is low relative to the prevailing macroeconomic conditions". Capturing the effect of a shock to financial conditions gives causal meaning to the model with controls, but the quantile in that model becomes conditional on controls. This may or may not be desirable, depending on the research agenda. Our framework allows to measure ``the effect of a shock to financial conditions when growth is low'' while still exploiting an identification by controls strategy. Maintaining this simpler interpretation is possible because we use the generalized quantile regression developed by \cite{Powell2020}, which explicitly distinguishes between treatment and control variables.

We identify the structural quantile impulse responses (QIRs) by combining generalized quantile regression of \cite{Powell2020} with local projections of \cite{Jorda2005} in a potential outcomes time-series framework similar to \cite{Angrist2011}. Our quantile impulse responses show the dynamic response of a chosen quantile of the distribution of the dependent variable to a shock to the treatment variable. This  definition makes comparisons with ubiquitous mean impulse responses straightforward. We achieve causal identification with timing restrictions in a controls-based strategy. As such, our identifying assumptions will look familiar to researchers who use local projections or vector autoregressions to identify structural impulse responses. Having said that, conditional uncorrelatedness which is sufficient to identify structural mean impulse responses has to be replaced with a stronger assumption of conditional independence to identify structural quantile impulse responses.

We motivate our framework by considering a stylized structural vector autoregression (SVAR) augmented by an endogenous stochastic volatility term. This relatively simple variation on the traditional SVAR, gives rise to quantile impulse responses that differ from the mean impulse response and vary across quantiles. It also generates skewed ergodic distributions of the model's endogenous variables, even though the structural shocks are Gaussian. We show that in this model the structural quantile function underpinning the quantile impulse responses is nonlinear in the shock. By using the model as a data generating process in a Monte Carlo experiment, we show that quantile local projections fail at recovering the true structural quantile function and thus also do not recover the structural quantile impulse response. The reason is not the presence of the non-linearity of the structural quantile function, but rather the fact that quantile local projections with controls are not estimating the structural quantile functions conditional on treatment only. Instead, the quantile regression of \cite{Koenker1978} used in quantile local projections estimates the slope coefficients of the conditional on treatment and controls quantiles of the outcome. In contrast, our local projections based method which uses the generalized quantile regression of \cite{Powell2020}, successfully recovers the desired structural quantile function and thus also identifies the structural quantile impulse responses.

We apply our framework to revisit the effect of financial shocks on industrial production growth in the US. We control for; macroeconomic, financial and monetary policy variables to recover the causal effect. Our methodology allows us to include these controls for identification while still modeling the unconditional quantiles of growth. Our findings show that financial shocks - whether pertaining to credit risk or volatility - cause large output losses, but only in low growth environments. The asymmetry in the effects of financial shocks across quantiles of the growth distribution is larger than previously thought \parencite{Adrian2019, Matthes2024}. We find persistent output losses of 2\% points from a one standard deviation credit risk shock for the low quantiles, with the median and high quantile losses of only 0.5\% points.

Various defintions of QIRs as well as methods of identifying and estimating them have been proposed in the literature. The quantile local projections framework in \cite{Ruzicka2021, Bochmann2023} is the closest to our approach. This framework recovers the QIR from local projection coefficients estimated using quantile regression of \cite{Koenker1978}, with estimation done separately for each quantile and horizon. In the absence of control variables, there is no difference between our framework and quantile local projections. However, when controls are included quantile local projections identify a conditional on treatment and controls QIR. \cite{Chavleishvili2023} achieve identification by imposing timing restrictions on a recursive quantile vector autoregressive model. They report QIRs which assume a realization of a median sample path for the shock variable over the response horiozon. \cite{Montes2019} and \cite{Lee2019} use the mean-based vector autoregression model to identify a structural shock since their multivariate quantile models are reduced-form. The QIR proposed by \cite{Montes2019} describes the cumulative impact of a series of shocks, not a one-off shock, because persistent realizations of lower (or upper) quantiles are assumed in its construction. \cite{Han2024} and \cite{Jung2020} study QIRs in models where the quantile itself is autoregressive, as in the CAViaR model of \cite{Engle2004}. In the applied literature, \cite{Mumtaz2015} study the heterogeneity in the transmission mechanism of monetary policy across stages of the business cycle. They estimate the structural QIRs using a quantile autoregressive-distributed lag model of \cite{Galvao2013} using lags of the monetary policy shock as observable structural shocks. \cite{Matthes2024} study the drivers of  macroeconomic tail risks using a two-stage method. In the first stage, they estimate a quantile regression and save the fitted quantiles of economic growth, which are then used in the second stage as a dependent variable in a conventional local projections framework.

This paper contributes to the existing literature by showing how to identify and estimate \textit{conditional on treatment only} QIRs in the presence of control variables. In addition, our framework combines other desirable properties. Structural identification is achieved in a single-stage procedure. Our QIRs show the impact of a one-off shock, making comparisons with mean impulse responses more direct. Our framework can be used to estimate non-linear QIRs which allow for the responses to depend on the shock size.

The remainder of the paper is structured as follows. Section \ref{sec:model} introduces the methodology and motivates it using a simulation study. Section \ref{sec:empirics} contains our empirical analysis. Concluding remarks follow in Section \ref{sec:end}.

\section{Econometric framework}\label{sec:model}

\subsection{Identification}

Our framework builds upon the literature on local projections \parencite{Jorda2005}, quantile treatment effects in a potential outcomes framework \parencite{Powell2020}, and potential outcomes for time series \parencite{Angrist2011}.

Let $\{ Y_{t} \}$ be an outcome variable of interest, let $\{ D_t \}$ bet a scalar treatment variable and let $\{ \mathbf{W_t} \}$ be a set of contemporaneous and past values of control variables.
We assume the joint process $\{ (Y_t , D_t , \mathbf{ W_t'} )' \}$ to be stationary. Throughout, we use capital letters to denote random variables and lower case letters to denote their realizations.

Following \cite{Angrist2011}, we define a potential outcome denoted ${Y_{t,h}(d)}$ as the value assumed by $Y_{t+h}$ if $D_t = d$, where d is a possible value of $D_t$. The set of potential outcomes includes the observed outcome $Y_{t+h} \equiv Y_{t,h}(D_t)$. ${Y_{t,h}(d)}$ is a random variable which depends on shocks up to $t+h$. Note that both the timing and the horizon of the treatment matter, for instance $Y_{t,h}(d)$ and ${Y_{t+1,h-1}(d)}$ may differ even though both occur in period $t +h$, because $Y_{t,h}(d)$ does not constrain the policy in period $t+1$ to equal $d$. For a fixed treatment $d$ and for each horizon $h$, we assume  ${Y_{t,h}(d)}$ has a  structural quantile function (SQF) denoted by $q_h(\tau \mid d)$. Note that the control variables do not enter into this SQF, which distinguishes it from the conditional on treatment and controls SQF denoted by $q_h(\tau \mid d, \mathbf{w'})$. Stationarity assures that the SQF exists and does not depend on time $t$.

In our framework, each latent outcome can be related to its quantile function  as follows:
\begin{equation*}
Y_{t,h}(d) = q_h( U_{t,h}(d) \mid d), U_{t,h}(d) \sim \mathsf{Uniform}[0,1].
\end{equation*}
$U_{t,h}(d)$ is responsible for heterogeneity of outcomes among time periods with the same treatment state $d$. We refer to it as a rank variable as it determines the placement in the h-periods ahead outcome distribution for a given treatment $d$. $U_{t,h}(d)$ contains information up to time $t+h$. 

The goal of this paper is to idenify the structural quantile impulse response, defined as:
\begin{equation}
\mathsf{QIR}_\tau =    \frac{\partial q_h(\tau \mid d)}{\partial d}.
\label{qir}
\end{equation}
Note that if the SQF is linear i.e.  $q_h(\tau \mid d) = \alpha_h(\tau) +\beta_h(\tau)d $, then the $\mathsf{QIR}_\tau = \beta_h(\tau)$ does not depend on $d$. We discuss whether linearity of the SQF can be justified later. Importantly, we contrast this definition with the structural conditional quantile impulse response defined as:
\begin{equation}
\mathsf{cQIR}_\tau =    \frac{\partial q_h(\tau \mid d, \mathbf{w'})}{\partial d}.
\label{cqir}
\end{equation}
The QIR and the cQIR may differ even if the treatment and control variables are independent ($D_t \indep \mathbf{W'_t}$). Furthermore, the same observation $Y_{t+h}$ might fall below $q_h(\tau \mid d, \mathbf{w'})$ but above $q_h(\tau \mid d)$ or vice versa.

If the observed treatment $D_t$ is randomly assigned i.e. $ U_{t,h}(d)\mid D_t \sim  U_{t,h}(d) \sim \mathsf{Uniform}[0,1] $, then a quantile local projection model $Y_{t+h} = q_h(U_{t+h} \mid D_t)$ estimated using a standard quantile regression restriction $P(Y_{t+h} \leq q_h(\tau \mid D_t) \mid D_t) = \tau$ identifies the QIR as defined in equation \ref{qir}. In non-experimental settings such as ours, an endogeneity problem arises because the realized treatment $D_t$ is not randomly assigned. We address the endogeneity problem with an identification by controls strategy. In particular, we relax the assumption that  $ U_{t,h}(d)\mid D_t \sim  U_{t,h}(d)$ and replace it with $ U_{t,h}(d)\mid D_t, \mathbf{W_t'} \sim  U_{t,h}(d) \mid \mathbf{W_t'}$.\footnote{Note that this allows for the rank variable to have different distributions for different values of the controls $\mathbf{W_t'}$. I.e. the controls can help predict whether the outcome will be below/above its conditional (on treatment) quantile.} In other words, we assume that the treatment is conditionally on (observable) controls randomly assigned. We think of the observed treatment as a function of the observable controls and an unobserved structural shock $Z^D_t$, i.e. $D_t = \delta( \mathbf{W_t'}, Z^D_t)$. As such the object of our causal analysis is the quantile impulse response to a structural shock to the treatment variable. 

The Frisch-Wough-Lovell theorem does not apply to quantile regression making disentangling effect of controls on identification versus interpretation more difficult. In particular, the quantile local projections model with controls $Y_{t+h} = q_h(U^*_{t+h} \mid D_t, \mathbf{W_t'})$ estimated using a restriction $P(Y_{t+h} \leq q_h(\tau \mid D_t , \mathbf{W_t'}) \mid D_t,  \mathbf{W_t'}) = \tau$  deals with the endogeneity issue, but estimates a different structural function $q_h(\tau \mid d, \mathbf{w'})$ instead of $q_h(\tau \mid d)$. As such it estimates the cQIR defined in equation \ref{cqir} instead of the QIR defined in equation \ref{qir}. The addition of controls into the equation changes the interpretation of the model. As such, even in cases when the treatment is randomly assigned, inclusion of control variables could change the quantile regression coefficients on the treatment variable.\footnote{For a simple example, consider a DGP given by $Y = DU + W$, where $D, U, W \overset{iid}\sim \mathsf{Uniform}[0,1]$.  The coefficients on $D$ from quantile regressions $Y = \alpha(U) + \beta(U) D$ and $Y = \alpha(U) + \gamma(U) D + \phi(U) W$ will differ, $\beta(\tau) \neq \gamma(\tau)$ unless $\tau = 0.5$.} Note also that the conditional on controls rank variable $U^*_{t,h}(d, \mathbf{w'})$ is distinct from $U_{t,h}(d)$. In particular, $U_{t,h}(d) = \lambda_{d,h} (\mathbf{W_t'},U^*_{t,h}(d, \mathbf{w'}))$ for some function $\lambda_{d,h}$ that depends on the fixed treatement and the horizon, but not time.

Exploiting control variables for causal identification while still modeling the conditional on treatment only SQF $q_h(\tau \mid d)$ is possible thanks to the \cite{Powell2020} Generalized Quantile Regression (GQR) framework, which explicitly distinguishes between treatment and control variables. We adapt this cross-sectional framework to our time-series setting and consider identification by controls only (\cite{Powell2020} also considers identification using instrumental variables). We assume the below assumptions hold for each $h \in \{0,1,2,...,H\}$:

\begin{assumption}[Potential Outcomes]
For a fixed $t$ and $h$, potential outcome $Y_{t,h}(d)$ is defined as the value that $Y_{t+h}$ would have taken had $D_t = d$ been observed. $Y_{t,h}(d)$ has a structural quantile function $q_h( \tau \mid d)$, where $\tau \mapsto q_h( \tau \mid d)$ is non-decreasing on $[0, 1]$ and left-continuous.
\label{as:1}
\end{assumption}
\begin{assumption}[Conditional Independence]
$Y_{t,h}(d) \mid D_t, \mathbf{W_t'} \sim Y_{t,h}(d) \mid \mathbf{W_t'}$. Potential outcomes $Y_{t,h}(d)$ are conditionally (on $\mathbf{W_t'}$) independent of the treatment $D_t$. 
\label{as:2}
\end{assumption}
\begin{assumption}[Rank Similarity]
$\mathbb{P}[Y_{t,h}(d) \leq q_h( \tau \mid d) \mid D_t, \mathbf{W_t'}]= \mathbb{P}[Y_{t,h}(d') \leq q_h(d', \tau)  \mid D_t, \mathbf{W_t'}] $, $\forall d, d'$. 
\label{as:3}
\end{assumption}
\begin{assumption}[Observability and Stationarity]
We observe $Y_{t+h}\equiv Y_{t,h}(D_t),D_t, \mathbf{W_t'}$.  $(Y_{t+h},D_t, \mathbf{W_t'})'$ is a jointly stationary time-series.
\label{as:4}
\end{assumption}

Before stating the moment conditions used to recover the QIR, we restate the Theorem 1 from \cite{Powell2020} except for our linear, time series setting. The proof of the theorem is in the appendix.
\begin{theorem}
Suppose Assumptions \ref{as:1}-\ref{as:4} hold, then $\forall h \in \{0,1,2,\dots, H\}$ and for each $\tau \in (0,1)$: \\
$ \mathbb{P}[Y_{t+h} \leq q_h( \tau \mid D_t) \mid  D_t,  \mathbf{W_t'}] = \mathbb{P}[Y_{t+h} \leq q_h( \tau \mid D_t)  \mid  \mathbf{W_t'}],$ \\
$\mathbb{P}[Y_{t+h} \leq  q_h( \tau \mid D_t)] = \tau.$
\label{th:powell1}
\end{theorem}

The first equation in theorem \ref{th:powell1}, states that once we condition on controls $\mathbf{W_t'}$, the treatment $D_t$ does not provide additional information about the probability that the outcome is below its quantile function. The second equation in theorem \ref{th:powell1}, ensures that the quantile function is correctly scaled. Together, these equations imply that the conditional probability $\mathbb{P}[Y_{t+h} \leq q_h( \tau \mid D_t)  \mid  \mathbf{W_t'}]$ is allowed to vary based on controls $\mathbf{W_t'}$, but in expectation it is equal to the quantile level $\tau$. When there are no control variables in the model (i.e. $ \mathbf{W_t'} = 0$), the two conditions in theorem  \ref{th:powell1} collapse into one standard quantile regression restriction $ \mathbb{P}[Y_{t+h} \leq q_h( \tau \mid D_t) \mid  D_t] = \tau$. This restriction is used to estimate QIRs in the quantile local projections framework. As such, quantile local projections are a special case in our framework, corresponding to a setting where all the model variables are treatment variables and there are no controls. Therefore, our framework ``nests'' the quantile local projections framework.

Theorem \ref{th:powell1} gives us two moment conditions for each $h \in \{0,1,2,\dots, H\}$: 
\begin{align*}
\mathbb{E}\{D_t[\mathbb{I} (Y_{t+h} \leq  q_h( \tau \mid D_t)) - \mathbb{P}(Y_{t+h} \leq  q_h( \tau \mid D_t) \mid \mathbf{W_t'}) ]\}=0, \\
\mathbb{E}[\mathbb{I} (Y_{t+h} \leq q_h( \tau \mid D_t)) - \tau ]=0,
\end{align*}
where $\mathbb{I}$ is the indicator function. For each quantile of interest $\tau$, estimation is done seperately for each horizon $h$ as in the local projections framework. For a given $h$ and $\tau$ and assuming a linear specification $q_h( \tau \mid d) = \alpha_h(\tau) + \beta_h(\tau) d$, estimation proceeds in three steps:
\begin{enumerate}
\item Postulate a candidate $\tilde{\beta}_h(\tau)$. For each candidate $\tilde{\beta}_h(\tau)$ there exists an intercept $\tilde{\alpha}_h(\tau)$ such that $\mathbb{P}(Y_{t+h}\leq \tilde{\alpha}_h(\tau) +\tilde{\beta}_h(\tau) D_t) = \tau$.
\item Given a pair of  $\tilde{\alpha}_h(\tau)$ and $\tilde{\beta}_h(\tau)$, estimate a binary outcome model (e.g. Logit or Probit) for the event that $Y_{t+h}\leq \tilde{\alpha}_h(\tau) + \tilde{\beta}_h(\tau) D_t$ as a function of controls $\mathbf{W'_t}$. Save the predicted probabilities as $\hat{\tau}_{\mathbf{W'_t}}$.
\item  $\hat{\beta}_h(\tau) = \operatorname*{argmin}_{\tilde{\beta}_h(\tau)} g' A g$, where $g = \frac{1}{T} \sum_{t=1}^{T} D_t[\mathbb{I} \{Y_{t+h} \leq \tilde{\alpha}_h(\tau)+ D_t \tilde{\beta}_h(\tau)\} -\hat{\tau}_{\mathbf{W'_t}} ]$.
\end{enumerate}
We use a grid-search algorithm to find the minimizer $\hat{\beta}_h(\tau)$, for more details about the estimation algorithm we refer the reader to \cite{Powell2020}.

We calculate confidence intervals using ``block-of-blocks" bootstrap. This procedure preserves the time-dependency by resampling blocks of $l$ consecutive $m$-tuples drawn from the set of all possible $m$-tuples \parencite{Kilian2017}. After re-estimating the model $B$ times using these pseudo-samples the confidence intervals are then based on the distribution of the estimated parameters across the $B$ replications of the procedure. 

\subsection{Connection with mean impulse responses}

When the treatment variable is continuous, the structural mean impulse response of $Y_t$ to a change in treatment $d$ can be defined as:
\begin{equation}
\mathsf{IR} = \frac{\partial   \mathbb{E}[Y_{t+h}\mid d]}{\partial d} .
\label{ir}
\end{equation}
Comparing the definition of the QIR in equation \ref{qir} and the definition of the IR in equation \ref{ir}, the similarity should be self-evident. The difference is that the QIR describes the dynamic causal effect of treatment on the quantile, rather than the expectation of the outcome variable. Both definitions capture the impact of a one-off shock to the value of the treatment at time $t$.  Overall, this makes comparisons with mean impulse responses more direct, compared to  alternative definitions of the quantile impulse response used in the literature.

\cite{Wolf2021} show that under appropriate assumptions the local projection and SVAR impulse responses are equal, up to a constant of proportionality. The presence of the constant of proportionality comes from the fact that the implicit local projection innovation (after controlling for the other right-hand side variables) does not have unit variance, unlike the innovations in a SVAR model. \cite{Wolf2021} provide an expression for this constant of proportionality, which makes it possible to compare the magnitude of the impulse responses estimated using local projection and SVAR frameworks. Similarly to the local projection mean impulse response, the QIR estimated by our model should be interpreted as a response to shock which causes a unit change in the treatment $D_t$, rather than a response to a unit shock (as is common in the SVAR literature). Therefore, if we want to compare the QIRs with mean impulse responses from a local projections, we can simply ignore the constant of proportionality, and if we want to compare them with impulse responses from a SVAR we scale the recovered impulse response by the constant of proportionality.

A word of caution is in order when dealing with cumulative quantile impulse responses. To calculate cumulative impact on growth in the level of the variable of interest (e.g. Industrial Production $IP_t$) using local projections, the outcome variable is usually transformed to $Y_{t+h} = log(IP_{t+h}) - log(IP_{t-1})$. This is also the transformation used in this paper. This transformation is innocuous in the case of the mean impulse response as linearity of the expectations operator implies:
\begin{equation*}
\sum_{s=0}^{h} \frac{\partial   \mathbb{E}[log(IP_{t+s}) - log(IP_{t+s-1})\mid d]}{\partial d} = \frac{\partial   \mathbb{E}[Y_{t+h}\mid d]}{\partial d}
\end{equation*}
meaning that the effect on average cumulative growth is equal to the sum of the effects on the consecutive between period average growth rates. Importantly, the cumulative log growth transformation $Y_{t+h} = log(IP_{t+h}) - log(IP_{t-1})$ is not as innocuous in the case of quantile impulse responses. This is because, a quantile of a sum random variables is generally not equal to the sum of the quantiles of the random variables, unless the random variables are comonotonic. For example, the effect on the median annual growth rate will not generally equal to the sum of the effects on the 12 consecutive median monthly growth rates. When $Y_{t+h}$ is cumulative growth the quantile impulse responses have the former interpretation. They describe how the $\tau$ quantile of the $h$ periods ahead distribution of cumulative growth is affected by a time $t$ change in treatment.

\subsection{Illustrative Example}

To illustrate our framework consider a Structural Autoregression (SVAR) augmented by an endogenous stochastic volatility term:
\begin{align*}
Y_t &= \rho_y Y_{t-1} + \delta_{dy} D_{t-1} + \frac{1 + \phi \sqrt{\exp(D_{t-1})}}{1+\phi} Z_t^Y \\
\gamma Y_t + D_t &=  \delta_{yd} Y_{t-1} + \rho_d D_{t-1} + Z_t^D
\end{align*}
Where, $Z_t^Y, Z_t^D \overset{iid}\sim N(0,1)$ are unobserved structural shocks.  If parameter $\phi = 0$ the model collapses to a standard SVAR with $Y_t$ ordered first ($Y_t$ predetermined with respect to $D_t$). When $\phi > 0$ the stochastic endogenous volatility term $\sqrt{\exp(X_{t-1})}$ creates a relationship between $D_{t-1}$ and the volatility of $Y_t$. This generates interesting volatility dynamics that give rise to a skewed ergodic distribution of $Y_t$ and QIRs that vary across quantiles. Note that the mean impulse responses in this model do not depend on the value of the volatility parameter $\phi$, they are the same as in the linear SVAR (case when $\phi = 0$). 

\begin{table}[h]
\centering
\begin{tabular}{l|llllll}
parameter & $\rho_y$ & $\rho_d$ & $\delta_{dy}$ & $\delta_{yd}$ & $\gamma$ & $\phi$ \\ \hline
value     & 0.5      & -0.1     & -0.25         & -0.1             & -0.2     & 9     
\end{tabular}
\caption{Model parameters used in the simulation.}
\end{table}

Although this is not an economic model,  to keep the discussion less abstract we think of $Y_t$ as economic growth and $D_t$ as the change in financial conditions (with postive values meaning tightening financial conditions). To study the cumulative impulse responses of the level of output we define a transformed dependent variable  $Y^c_{t+h} \equiv \sum_{j=0}^h Y_{t+j} $. 

The structural mean impulse responses can be identified using local projections with appropriate timing restrictions \parencite{Jorda2005, Wolf2021}. Estimating the below by least squares separately for each $h\in \{1,2, \dots, H\}$:
\begin{equation*}
Y^c_{t+h} = \alpha_h +  D_t \beta_h + \mathbf{W_t'} \mathbf{\theta_h}  + \varepsilon_{t+h}
\end{equation*}
with $\mathbf{W_t'} = (Y_t, D_{t-1}, Y_{t-1})$, $\beta_h$ recovers the structural mean impulse response. We know which variables need to be included in $\mathbf{W_t'}$ from looking at the equation for $D_t$ in our data generating process and using the fact that $D_t = \delta( \mathbf{W_t'}, Z^D_t)$. The inclusion of the controls vector $\mathbf{W_t'}$ is necessary as $D_t$ is an endogenous variable, not including the correct variables in $\mathbf{W_t'}$ would result in biased estimates of the impulse response. If we could observe the structural shock $Z^D_t$ directly, we could replace $D_t$ with $Z^D_t$ and the controls $\mathbf{W_t'}$ would no longer be needed for $\beta_h$ to identify the structural impulse response (although their inclusion may still be desirable to improve the precision of the estimates). 

When interest lies in identifying the QIR as defined in equation \ref{qir}, employing the \cite{Koenker1978} estimator in a local projections setting might not be enough. Firstly, a linear quantile regression may be misspecified as we have not shown that the functional form of SQF $q_h(\tau \mid d)$ is linear. In short timeseries typical in macroeconomics, non-parametric estimation of the SQF may be unfeasible, especially for more extremes quantiles.  For any given model for the underlying data generating process we can try to characterize the implied functional form of $q_h( \tau \mid d)$. Depending on the model, a closed-form solution for the SQF may be hard to find from the model's equations. For example, a linear SVAR model (case when $\phi = 0$) has linear SQFs of endogenous variables to structural shocks. Furthermore, SVAR quantile impulse responses equal to the mean impulse response for all quantiles.\footnote{This is because shocks in a linear SVAR only have location shifting effects on the distribution of endogenous variables.} On the other hand, our augmented SVAR ($\phi > 0 $) which features non-trivial QIRs   -- ones that change depending on chosen quantile $\tau$ -- also features non-linear SQFs for some quantiles and horizons. Even if the model implied SQF might be hard to characterize in closed-form, for a  given model we can recover the shape of the SQF by simulation.

\begin{figure}[h!]
\centering
    \begin{subfigure}[b]{0.49\textwidth}
        \centering
        \includegraphics[width=\textwidth]{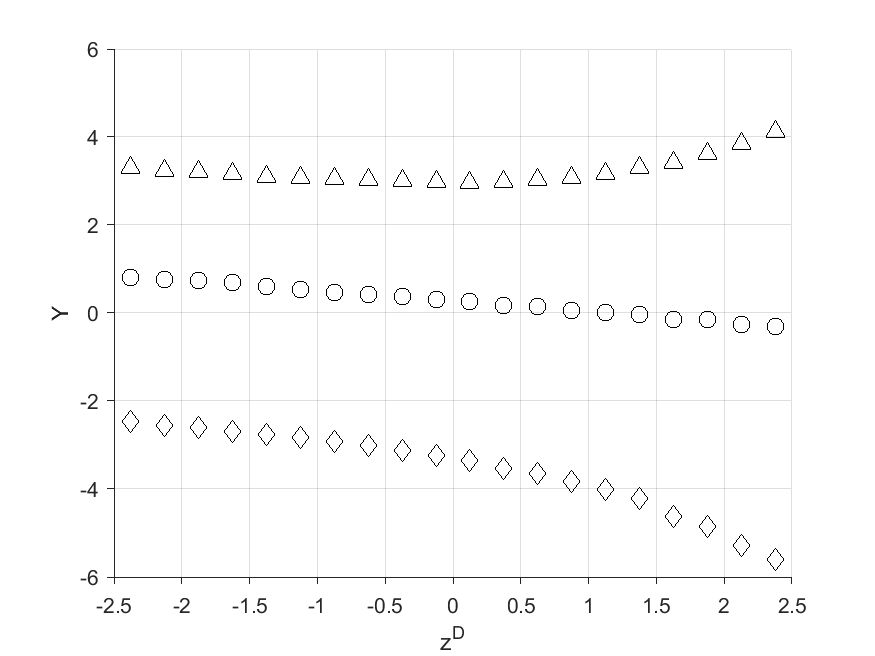}
        \caption*{True SQF}
    \end{subfigure}
    \begin{subfigure}[b]{0.49\textwidth}
        \centering
        \includegraphics[width=\textwidth]{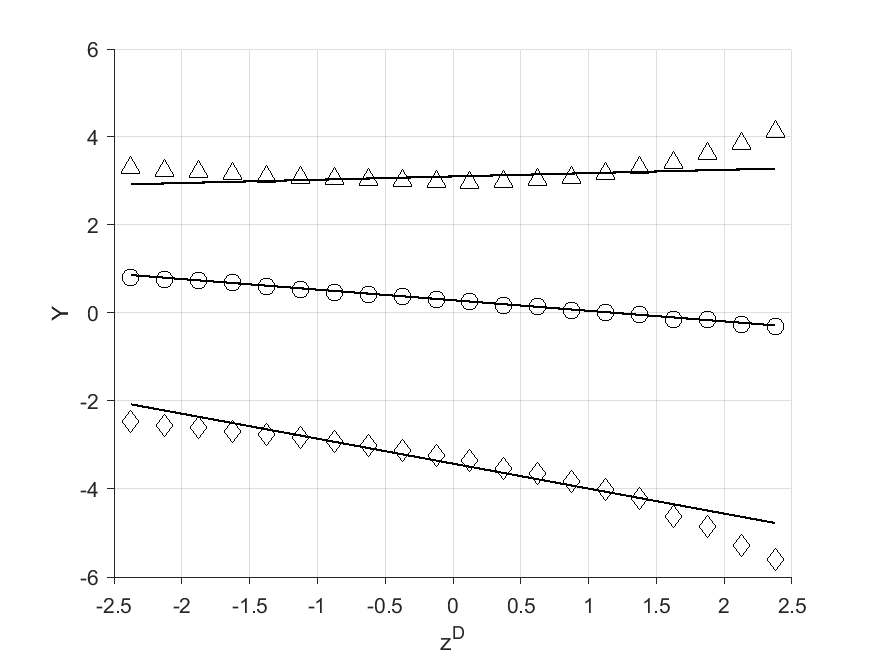}
        \caption*{Fitted linear SQF using QR}
    \end{subfigure}

    \begin{subfigure}[b]{0.49\textwidth}
        \centering
        \includegraphics[width=\textwidth]{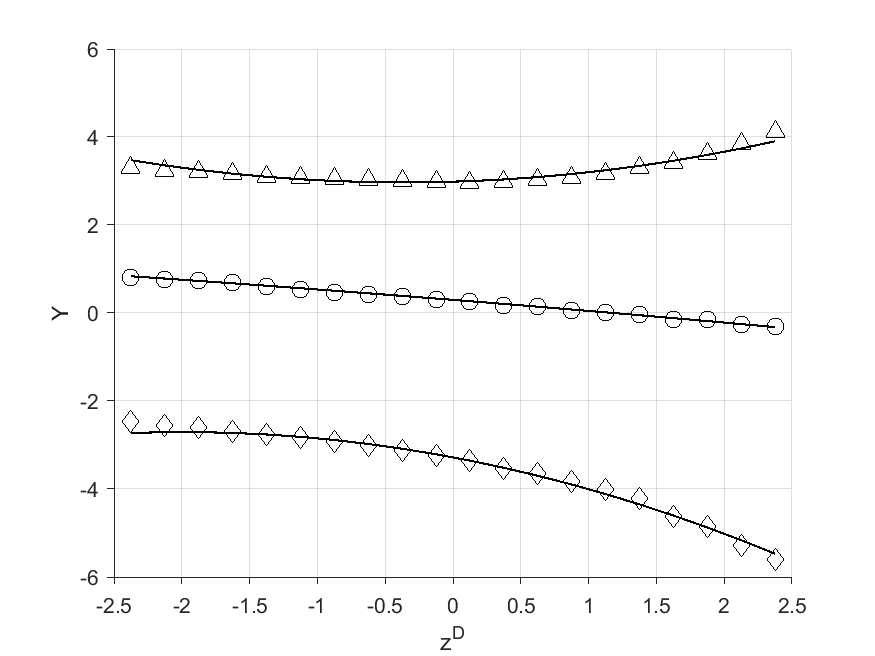}
        \caption*{Fitted quadratic SQF using QR}
    \end{subfigure}
	\caption{Simulation results for the first horizon SQF $q_1(\tau \mid d)$ of our illustrative example. The top-left panel plots the simulated quantiles of $Y^c_{t+1}$ over bins of $Z_t^D$ for quantiles $\tau \in \{ 0.1 \diamond, 0.5 \circ ,0.9 \scalebox{0.6}{$\triangle$} \}$ (obtained from a single simulated time-series of length $T = 1,000,000$). The other two panels re-plot these simulated quantiles, with the overlayed solid lines showing the fitted SQF using a quantile regression of $Y^c_{t+1}$ on the structural shock $Z_t^D$ for the same three quantiles. The fit in the top-right panel comes from a linear quantile regression while the bottom panel fit comes from a quadratic quantile regression. The regression coefficients used to plot the fitted SQFs are averaged estimates from a monte carlo simulation with $MC = 100$ replications and time-series of length $T = 1,000$ (after dropping $1,000$ initial observations).}
    \captionsetup{labelformat=empty}
    \label{fig:SQF}
\end{figure}

For a given set of parameters, we can simulate the model and plot the empirical quantiles of $Y^c_{t+h}$ over bins of $Z_t^D$ to recover the shape of the SQF. The top-left panel of figure \ref{fig:SQF} shows the first horizon SQFs for three quantiles $\tau \in \{0.1,0.5,0.9\}$, recovered from simulations using this method. Visual examination suggests that the SQF is quadratic for quantiles $\tau \in \{0.1,0.9\}$ and linear for the median $\tau = 0.5$. In the simulation setting we observe the structural shock $Z_t^D$ which is by construction independent. This suggest another strategy to recover the true SQF by estimating a quantile local projection model $Y^c_{t+h} = q_h(U_{t+h} \mid Z^D_t)$. Again, this requires either the knowledge of the functional form of the SQF, or the use of some non-parametric method to approximate it. Alternatively, we could (incorrectly) assume a linear specification, which although misspecified may nevertheless be a good approximation to the truth \parencite{Angrist2006}. Figure \ref{fig:SQF} shows that a linear quantile regression $Y^c_{t+1} = \alpha(U_{t+1}) + \beta(U_{t+1}) Z^D_t$ does well at approximating the true SQF around $Z^D_t = 0$, but is outperformed by a quadratic specification.

\begin{figure}[h!]
    \centering
    \begin{subfigure}[b]{0.49\textwidth}
        \centering
        \includegraphics[width=\textwidth]{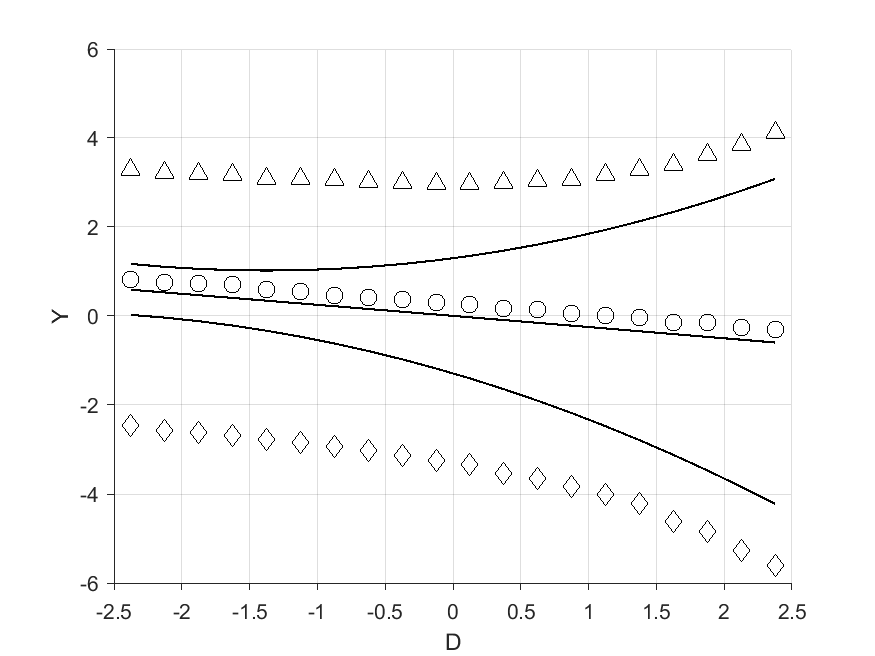}
        \caption*{QR no controls}
    \end{subfigure}
    \begin{subfigure}[b]{0.49\textwidth}
        \centering
        \includegraphics[width=\textwidth]{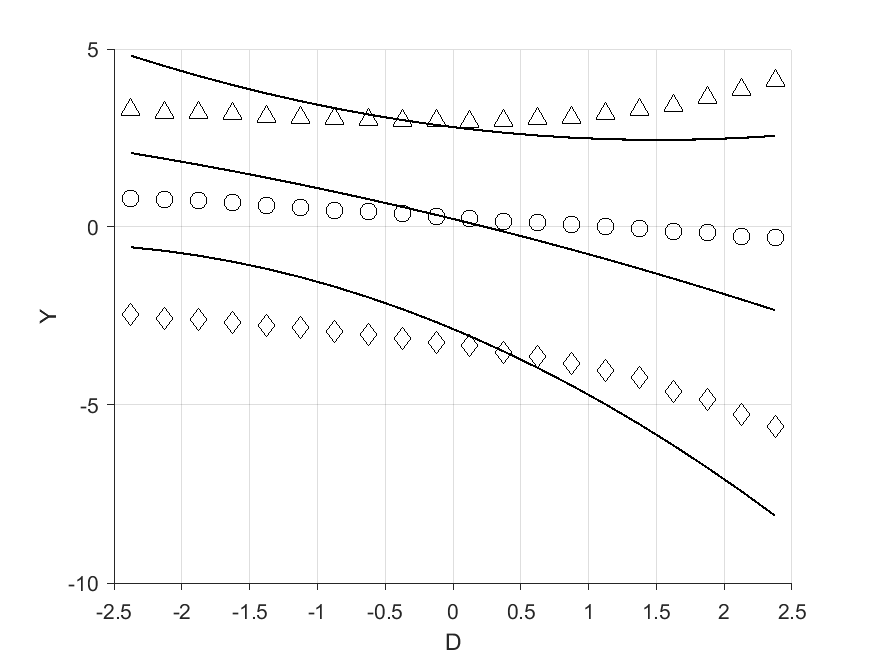}
        \caption*{QR with controls}
    \end{subfigure}

    \begin{subfigure}[b]{0.49\textwidth}
        \centering
        \includegraphics[width=\textwidth]{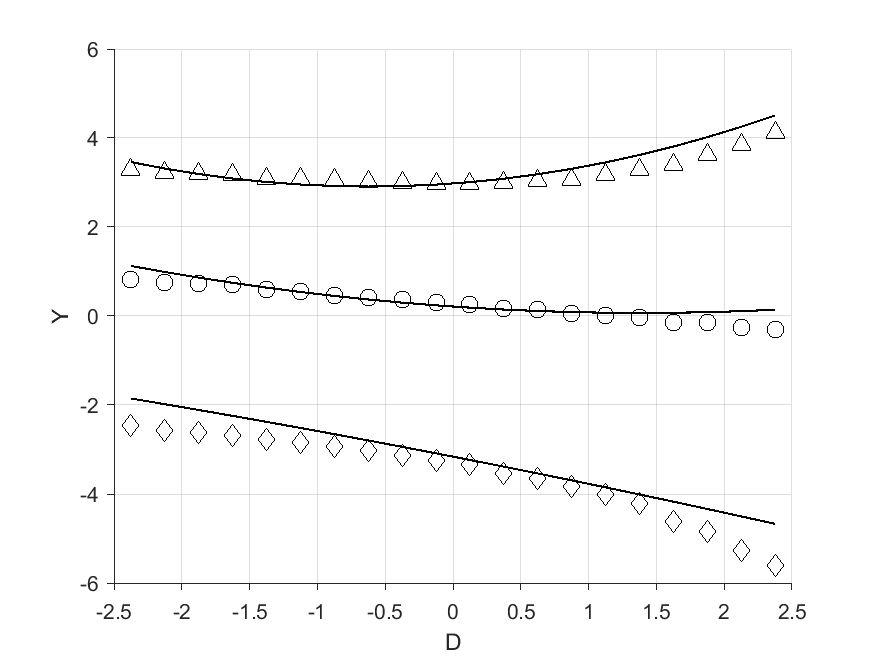}
        \caption*{GQR}
    \end{subfigure}
	\caption{Simulation results for the first horizon SQF $q_1(\tau \mid d)$ of our illustrative example. The diamonds, circles and triangles are the same across the three panels and show the simulated quantiles of $Y^c_{t+1}$ over bins of $Z_t^D$ for quantiles $\tau \in \{ 0.1 \diamond, 0.5 \circ ,0.9 \scalebox{0.6}{$\triangle$} \}$ (obtained from a single simulated time-series with T = 1,000,000). The three panels compare the performance of three estimators for the first horizon SQF. QR refers to the \cite{Koenker1978} estimator, GQR is the generalized quantile regression estimator introduced  by \cite{Powell2020}. The regression coefficients used to plot the fitted SQFs are averaged estimates from a monte carlo simulation with $MC = 100$ replications and time-series of length $T=1,000$ (after dropping $1,000$ initial observations).}
    \captionsetup{labelformat=empty}
    \label{fig:MC_quadratic}
\end{figure}

Outside of simulation settings we do not observe the structural shock $Z^D_t$, so we need to rely on the time-series of the endogenous model variables $\{Y_t, X_t\}$ to estimate the SQF. Figure \ref{fig:MC_quadratic} compares the performance of three quadratic models for the estimation of the true SQF at horizon $h=1$. The first is a model without controls estimated using quantile regression given by:
\begin{equation*}
Y^c_{t+1} = \alpha_{h=1}(U_{t+1}) + \beta_{1,h=1}(U_{t+1}) D_t+ \beta_{2,h=1}(U_{t+1}) D^2_t. 
\end{equation*}
The second model adds controls $\mathbf{W_t'}$ into the estimation equation and uses the quantile regression to estimate the parameters. 
\begin{equation*}
Y^c_{t+1} = \alpha_{h=1}(U_{t+1}) + \beta_{1,h=1}(U_{t+1}) D_t+ \beta_{2,h=1}(U_{t+1}) D^2_t + \mathbf{W_t'} \mathbf{\theta_{h=1}(U_{t+1})}. 
\end{equation*}
The third model also estimates the quadratic equation, but it uses the \textcite{Powell2020} generalized quantile regression which uses the controls  $\mathbf{W_t'}$ for identification, while modelling the quadratic SQF that is not conditional on controls.

Comparing the performance of these three models in recovering the SQF shows that the standard quantile regression is unable to recover the true shape of the SQF. The quantile regression model without controls suffers from endogeneity bias, while the quantile regression model with controls estimates a conditional SQF. This is why we opt for the generalized quantile regression estimator which targets the correct (unconditional on controls) SQF but is able to address the endogeneity of the treatment.

Since non-linear SQFs imply that the QIRs will vary not only with the quantile but also with the shock size, they make plotting and analyzing the QIRs more complicated. Thus for the sake of simplicity, we may decide to use the linear model even if we believe it to be misspecified. Ignoring the non-linearity of the SQF for the moment, we can recover (an approximation to) the QIR as the $\beta_h$ from the quantile local projection: 
\begin{equation*}
Y^c_{t+h} = \alpha_h(U_{t+h}) + \beta_h(U_{t+h}) Z^D_t.
\end{equation*}
This simple strategy is possible in a simulation setting as we can observe the structural shock $Z^D_t$, which by construction is independent ($U_{t+h}\mid Z^D_t \sim U_{t+h}$)  and thus controls are not needed for identification. A reasonable goal for an estimator of a structural QIR would be to recover the same QIR using only the time-series of the observed endogenous model variables $\{Y_t, D_t\}$, similarly to how we can recover the structural mean impulse response using local projections with the correct controls. Figure \ref{fig:MC} shows that quantile regression fails at achieving this goal.\footnote{In the appendix, we provide a table that compares the mean bias and root mean squared error of the three estimators up to horizon 10, to complement figure \ref{fig:MC}.} In particular, a quantile local projection model without controls:
\begin{equation*}
Y^c_{t+h} = \alpha_h(U_{t+h}) + \beta_h(U_{t+h}) D_t,
\end{equation*}
suffers from endogeneity of $D_t$ and as expected it fails to recover the structural QIR. Perhaps more surprisingly, a quantile local projection with the correct controls $\mathbf{W_t'} = (Y_t, D_{t-1}, Y_{t-1})$, given by:
\begin{equation*}
Y^c_{t+h} = \alpha_h(U_{t+h}) + \beta_h(U_{t+h}) D_t  + \mathbf{W_t'} \mathbf{\theta_h(U_{t+h})},
\end{equation*}
solves the endogeneity of $D_t$ problem, but in doing so models a conditional on treatment and controls SQF which has a different meaning than the conditional on treatment only SQF. In effect, it recovers the cQIR rather than the QIR, which in this case are not equal.

Exploiting the \cite{Powell2020} generalized quantile regression estimator in a local projection setting allows us to keep modelling the conditional on treatment only SQF, while still addressing the endogeneity of $D_t$. The GQR estimator with the dependent variable $Y^c_{t+h}$, treatment variable $D_t$ and controls $\mathbf{W_t'} = (Y_t, D_{t-1}, Y_{t-1})$ recovers the same QIRs as the (unfeasible in practice) QLP of $Y^c_{t+h}$ on the structural shock $Z^D_t$.

\section{Empirical Results}\label{sec:empirics}

\subsection{Data}

Our monthly dataset covers the US economy during the period between January 1985 and August 2023 (T=464). All of the data we use in the paper is publicly available, with majority of it contained in the FRED-MD database published by the St. Luis Fed \parencite{mccracken2015}. We use monthly data to benefit from a larger sample size. Due to the unavailability of monthly GDP we focus on Industrial Production (IP) as the dependent variable. This is a natural choice, as IP accounts for the bulk of the variation in output over the course of the business cycle. 

Throughout, the dependent variable $Y_{t+h}$ will be defined as the $h$-months cumulative log growth rate $Y_{t+h} =100*[ log(IP_{t+h})-log(IP_{t-1})]$. We multiply the log growth rates by 100 to interpret the QIR in terms of percentage points. We Z-score normalize the treatment variable $D_t$ to interpret the QIRs as responses to a one standard deviation change.

The first treatment variable $D_t$ we consider measures movements in credit risk. We will refer to this variable as credit risk and we define it as the first difference of the monthly Excess Bond Premium (EBP) of \cite{Gilchrist2012}, i.e. $D_t = EBP_{t}-EBP_{t-1}$. The EBP is a residual of corporate bond credit spreads that cannot be explained by movements in expected default risk, as such it measures the investor sentiment or risk appetite in the corporate bond market. 

The second treatment variable $D_t$ we consider measures the volatility risk premium in the equity markets, defined it as the difference between realized and implied volatility of the S\&P500 index. We will refer to it as volatility risk for short. If option markets are efficient, implied volatility should be an efficient forecast of future volatility, it should subsume the information contained in all other variables in the market information set in explaining future volatility. Thus, $D_t = realized_t - implied_t$  captures realized volatility that was unexpected by the financial markets \parencite{Christensen1998}.

We follow \cite{Gilchrist2012} in ordering the financial risk variable after macroeconomic variables but before financial markets and monetary policy variables. Our variables are ordered as follows: \{consumption growth, investment growth, industrial production growth, inflation, financial risk variable $D_t$, S\&P500 monthly return, change in the ten-year (nominal) Treasury yield, change in the effective (nominal) federal funds rate\}. This ordering implies that $\mathbf{W'_t}$ must include the contemporaneous values of the four variables ordered before the treatment variable $D_t$. Additionally, to control for the broad state of the economy in the recent past, we include the first two lags of all eight variables contained in our ordering in $\mathbf{W'_t}$. In short, our timing restriction assumption allows for financial conditions to adjust within' the period to consumption growth, investment growth, industrial production growth and inflation, but not to the stock market return, changes of the Treasury yields and changes to the Fed’s funds rate.

\subsection{Results}

Throughout, we focus on three quantiles $\tau \in \{0.1, 0.5, 0.9\}$. The $\tau = 0.1$ quantile is of primary interest as it represents downside-risk.\footnote{In the appendix, we also report results for a richer set of quantiles for four selected horizons, including the $\tau = 0.05$ quantile corresponding to the usual definition of GaR.} To simplify the analysis we assume a linear specification for the SQF $q_h(\tau \mid d) = \alpha_h(\tau) + \beta_h(\tau)d$, this ensures that the $\mathsf{QIR}_\tau  = \beta_h(\tau)$ does not depend on $d$. We choose a 90\% confidence level for reporting our block-of-blocks bootstrap confidence intervals, which are computed using a block length of $7$ and $1,000$ bootstrap replications.

\begin{figure}[h!]
    \centering
    \begin{subfigure}[b]{0.49\textwidth}
        \includegraphics[width=\textwidth]{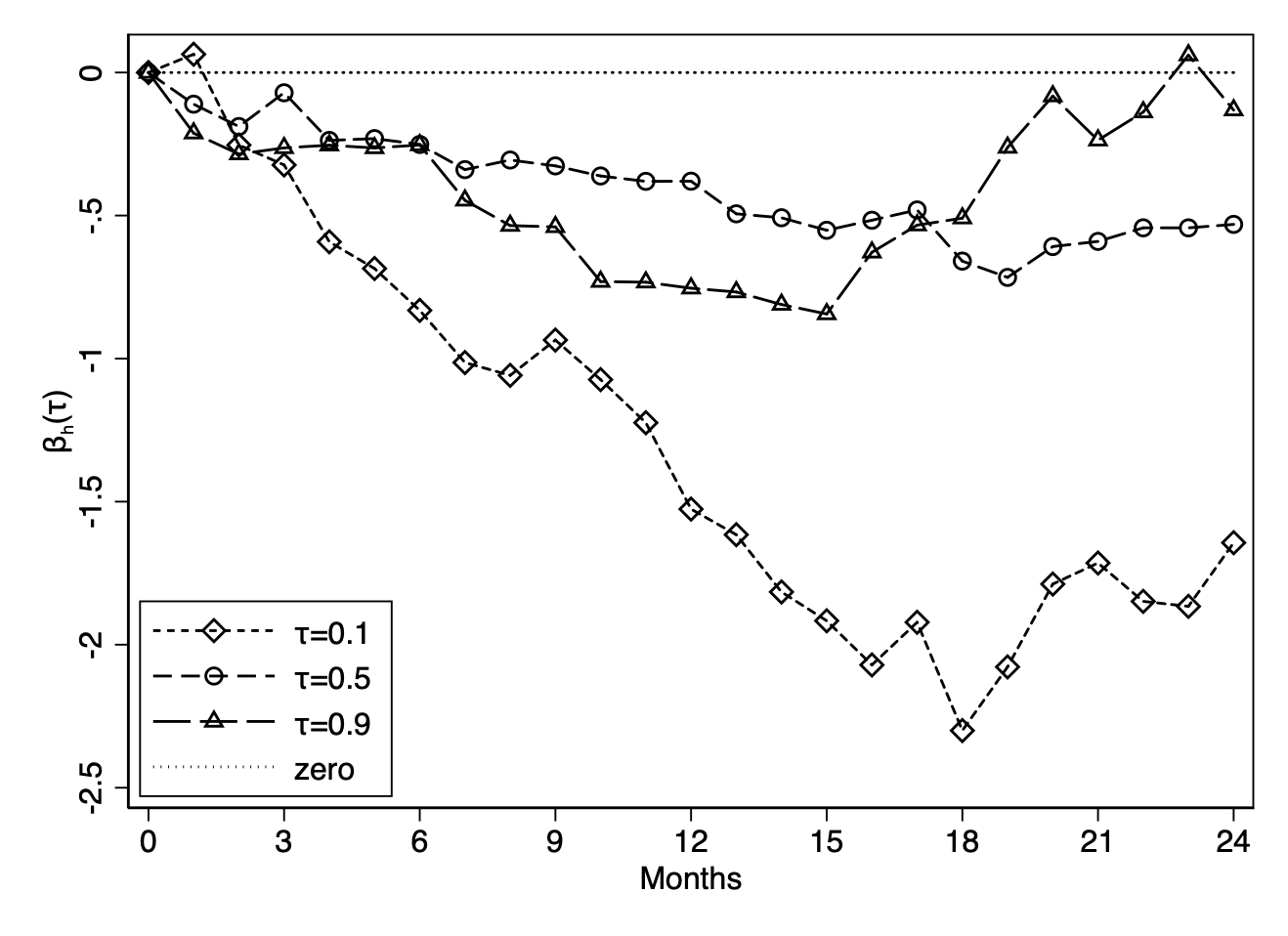}
	\caption{ $\tau \in  \{ 0.1 , 0.5, 0.9 \}$}
    \end{subfigure}
        \hfill
        \begin{subfigure}[b]{0.49\textwidth}
        \includegraphics[width=\textwidth]{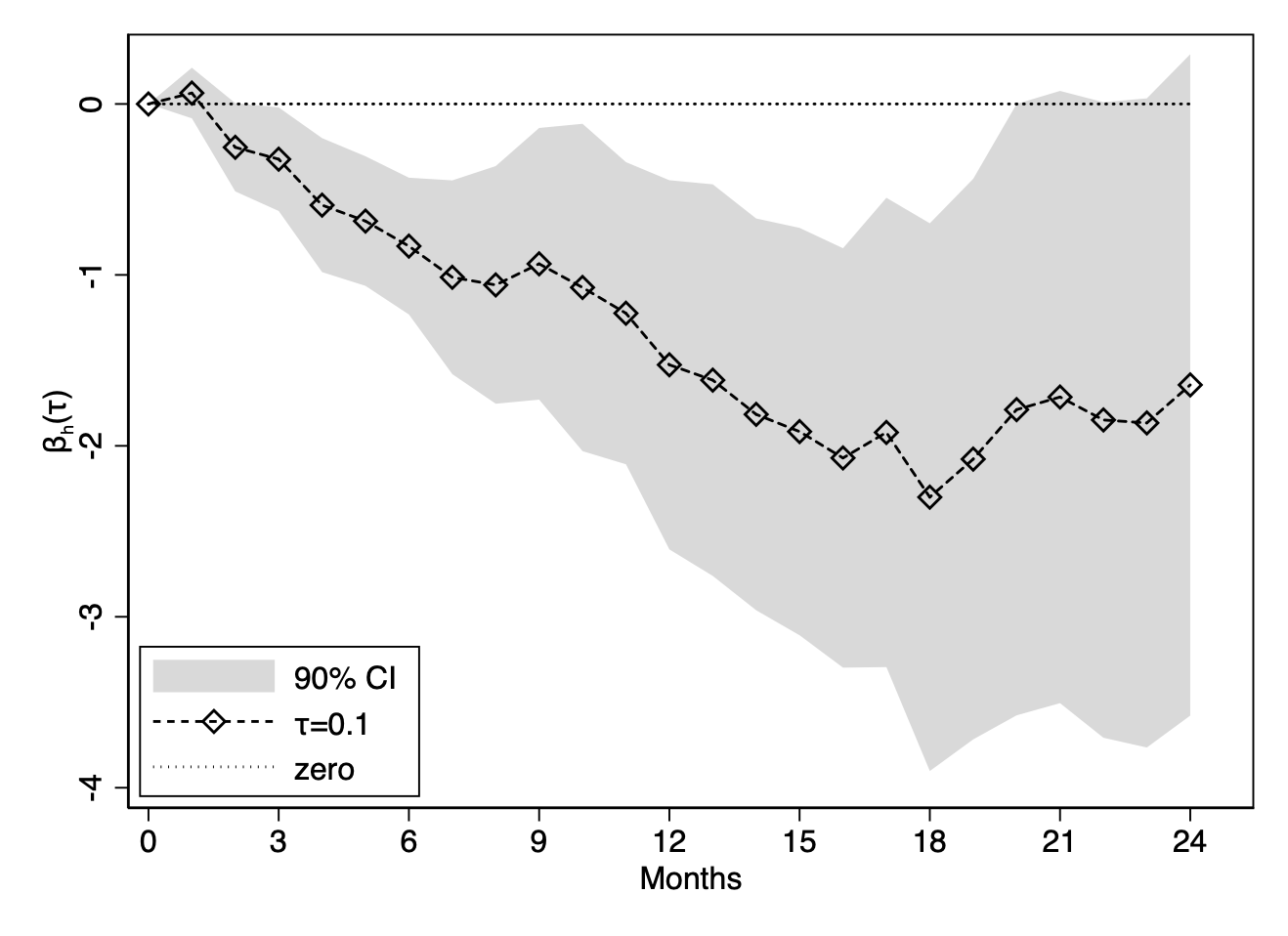}
	    \caption{ $\tau = 0.1$ with 90\% C.I.}
    \end{subfigure}

    \begin{subfigure}[b]{0.49\textwidth}
        \includegraphics[width=\textwidth]{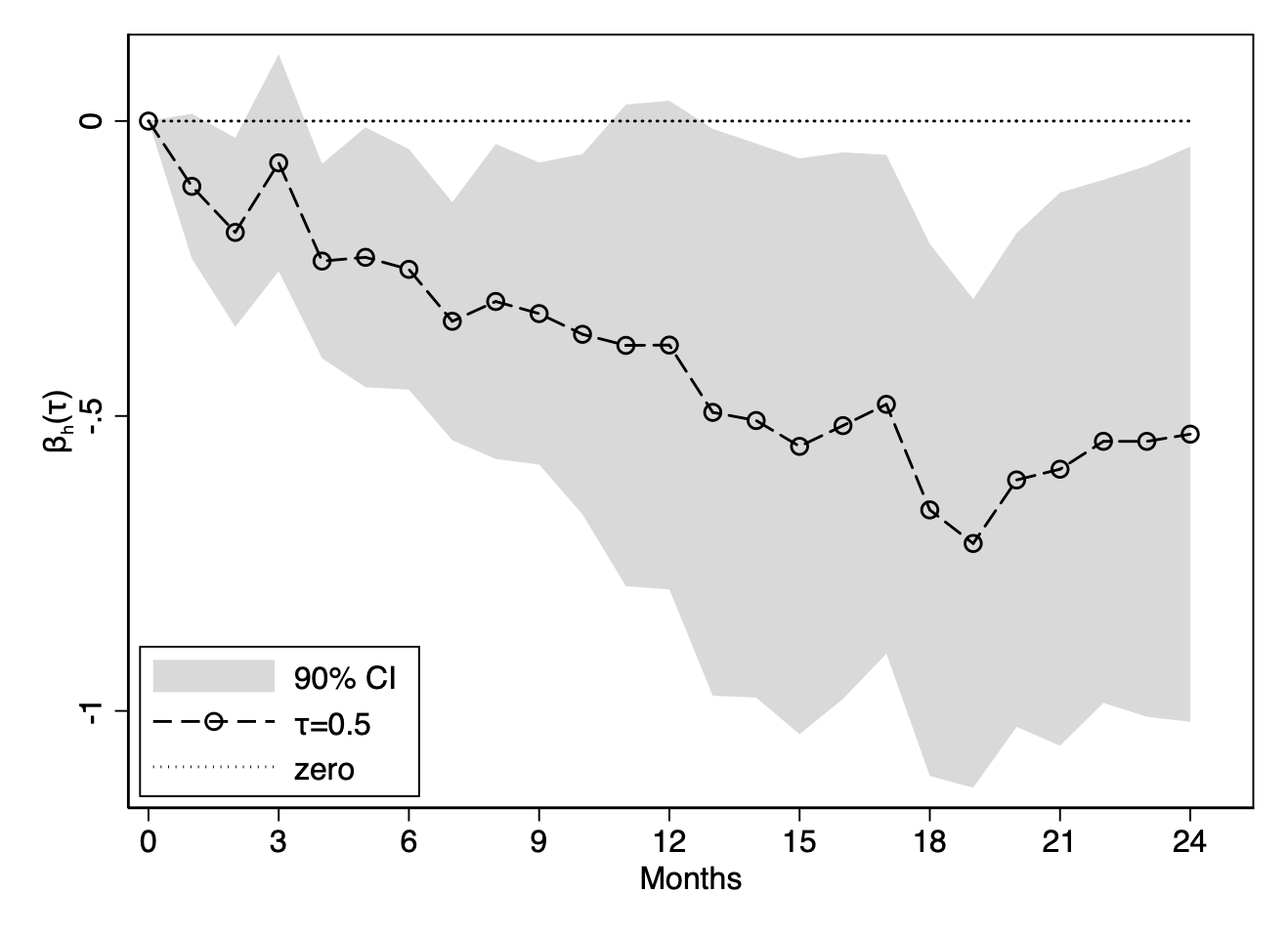}
	\caption{ $\tau = 0.5$ with 90\% C.I.}
    \end{subfigure}
        \hfill
        \begin{subfigure}[b]{0.49\textwidth}
        \includegraphics[width=\textwidth]{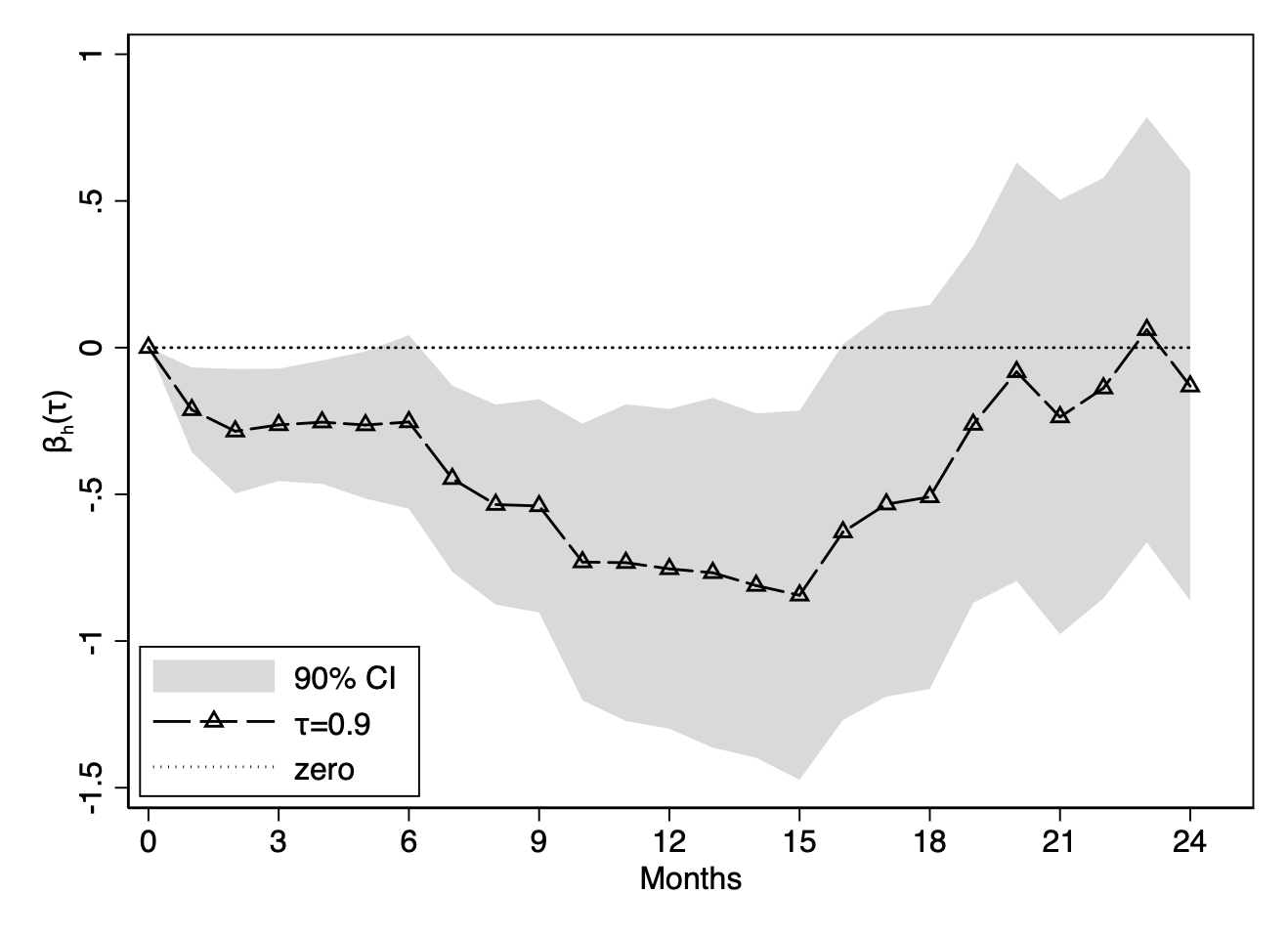}
	    \caption{ $\tau = 0.9$ with 90\% C.I.}
    \end{subfigure}
       \caption{Cumulative response of Industrial Production (in \% pts.) from a shock that increases credit risk by one standard deviation, plotted for three quantiles $\tau \in \{0.1,0.5,0.9\}$. Y-axis is the estimated response $\hat{\beta}_h(\tau)$, x-axis is the horizon $h$ in months. Dashed lines plot the quantile impulse response. Shaded area is the block-of-block bootstrap 90\% Confidence Interval (with block length of 7, and 1,000 bootstrap replications). Note that the impact response (horizon $h=0$) is by assumption zero, given our timing restrictions.}
\label{QIR_ebp}
\end{figure}

Figure \ref{QIR_ebp} shows the recovered QIRs of industrial production to a shock which increases credit risk by one standard deviation. The upper-left panel in figure \ref{QIR_ebp} plots the QIRs for the three quantiles on the same axis. It is clear that the response at the $\tau=0.1$ quantile is much more pronounced than the response at the other quantiles considered. This is a feature of the data and not of the model, as nothing is restricting the responses of lower quantiles to be lower than those of the upper quantiles.\footnote{For instance, a shock that lowers the variance of a distribution would give rise to positive QIRs for quantiles below the median and negative QIRs for quantiles above the median.}

Our findings suggest economically large and statistically significant (at 90\% confidence level) growth losses of about 2\% points when a credit risk shock propagates in a low growth environment ($\tau = 0.1$). The median losses ($\tau = 0.5$) are considerably smaller at around 0.5\% points. Reassuringly, our median impulse response is of similar magnitude as the impulse response of real GDP to EBP shocks estimated by \cite{Gilchrist2012} using a SVAR model with quarterly data. The upside-risk response ($\tau = 0.9$) is similar to the median response, except that the effect is not statistically significant beyond the fifteen months horizon. The estimation uncertainty measured by the block-of-blocks bootstrap confidence intervals increases with the horizon, it is also higher for the $\tau=0.1$ quantile than the median and the $\tau=0.9$ quantile.

\begin{figure}[h!]
    \centering
    \begin{subfigure}[b]{0.49\textwidth}
        \includegraphics[width=\textwidth]{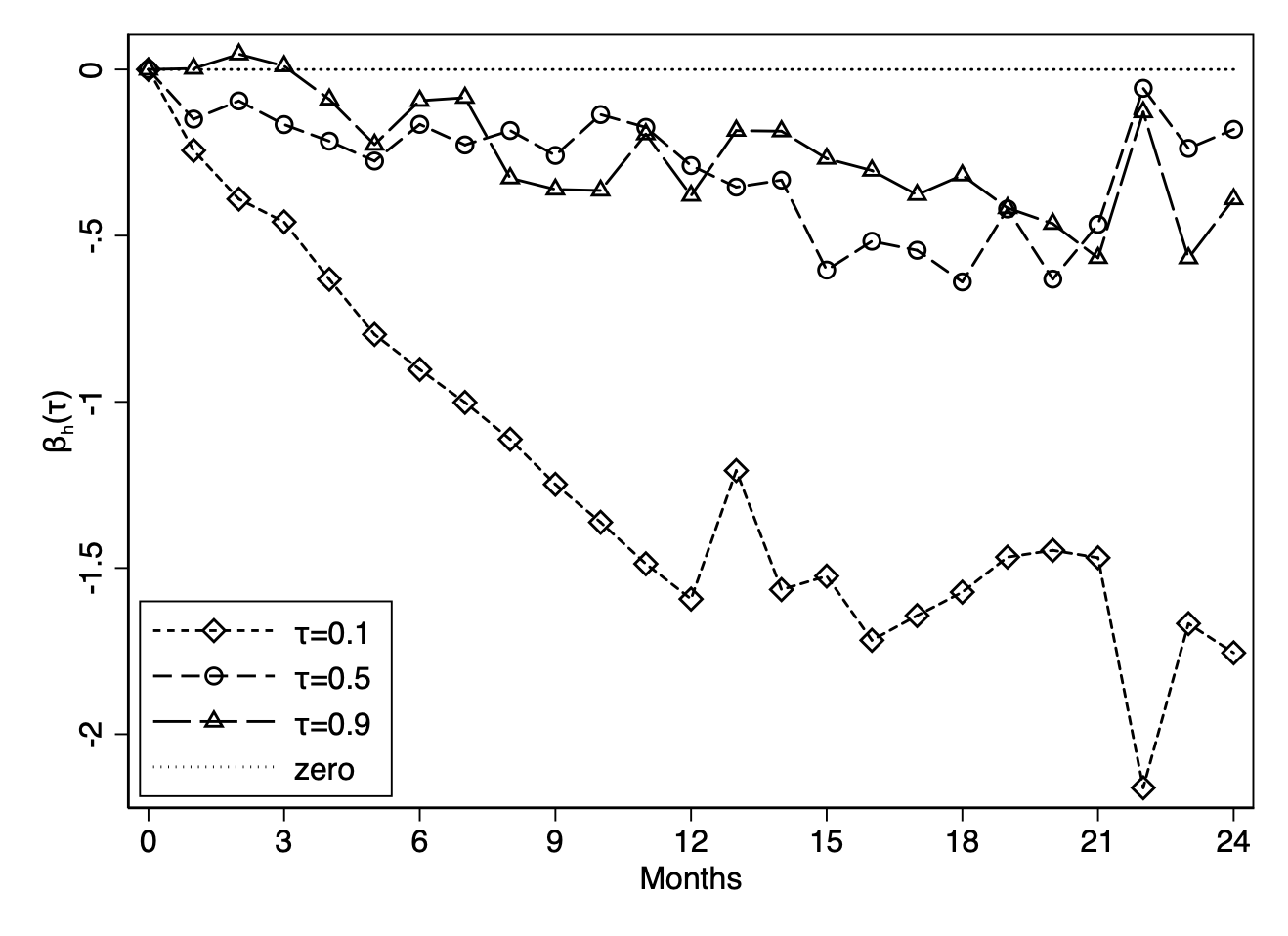}
	\caption{ $\tau \in  \{ 0.1, 0.5, 0.9\}$}
    \end{subfigure}
        \begin{subfigure}[b]{0.49\textwidth}
        \includegraphics[width=\textwidth]{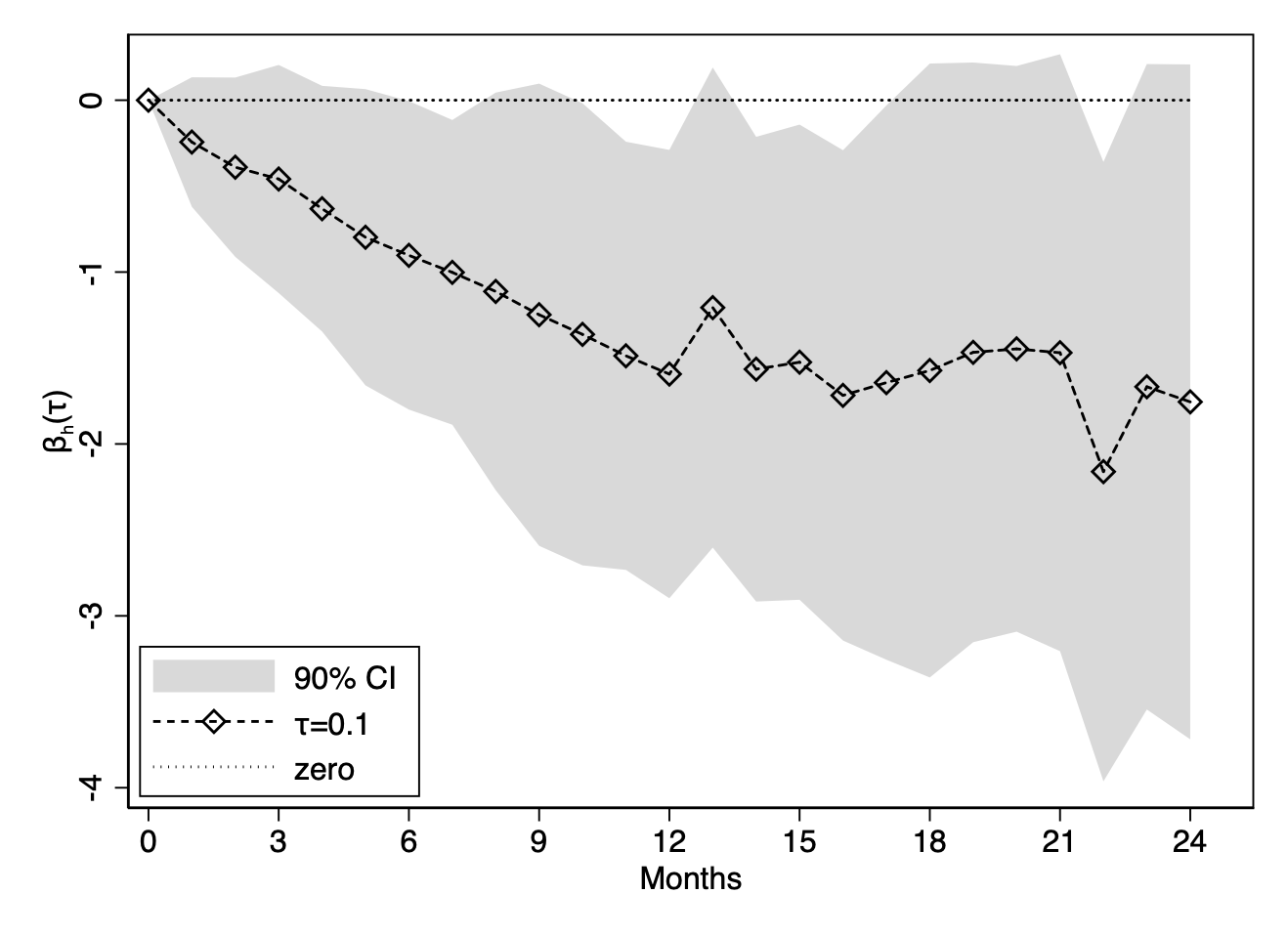}
	    \caption{ $\tau = 0.1$ with 90\% C.I.}
    \end{subfigure}

    \begin{subfigure}[b]{0.49\textwidth}
        \includegraphics[width=\textwidth]{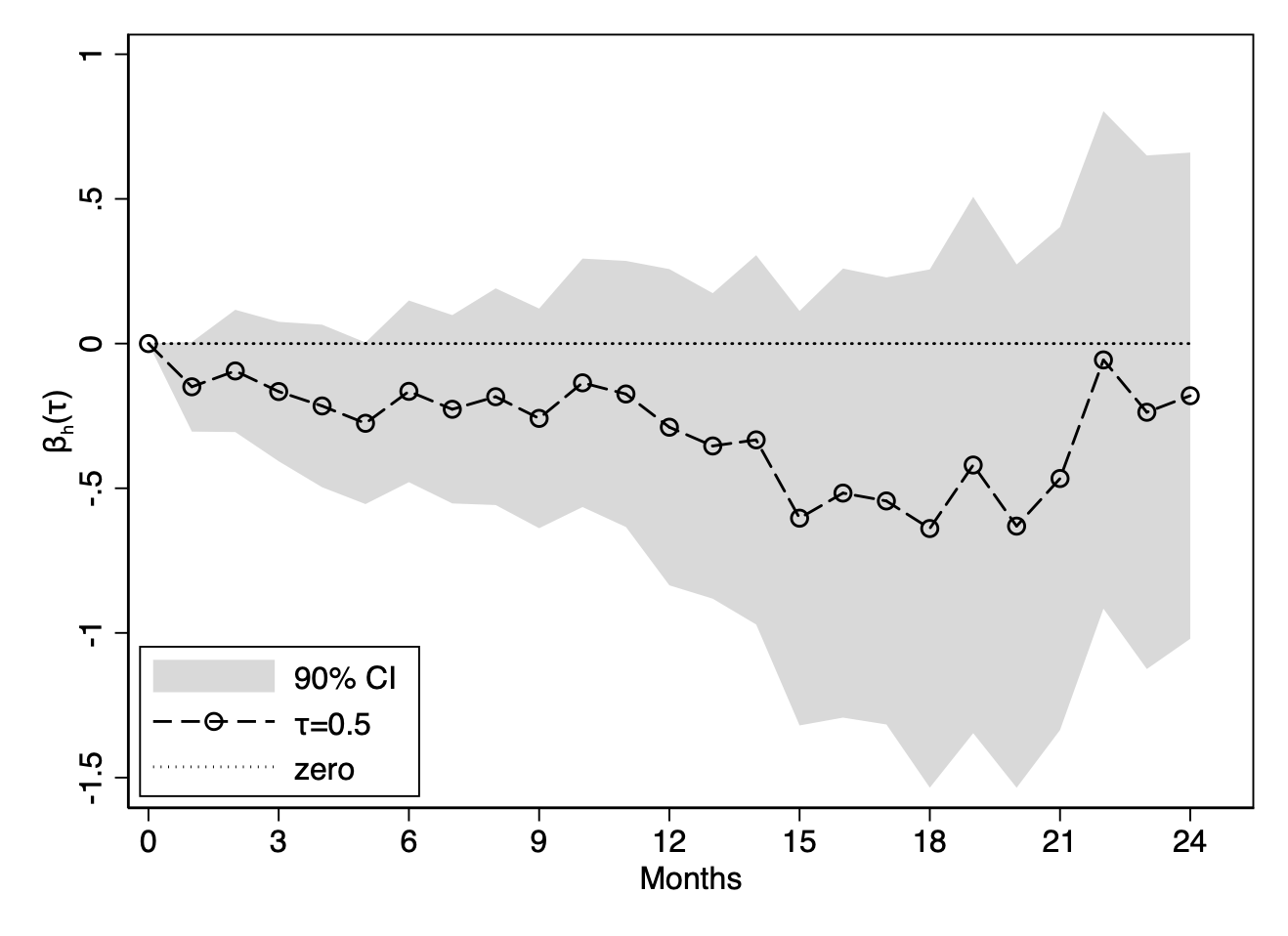}
	\caption{ $\tau = 0.5$ with 90\% C.I.}
    \end{subfigure}
        \hfill
        \begin{subfigure}[b]{0.49\textwidth}
        \includegraphics[width=\textwidth]{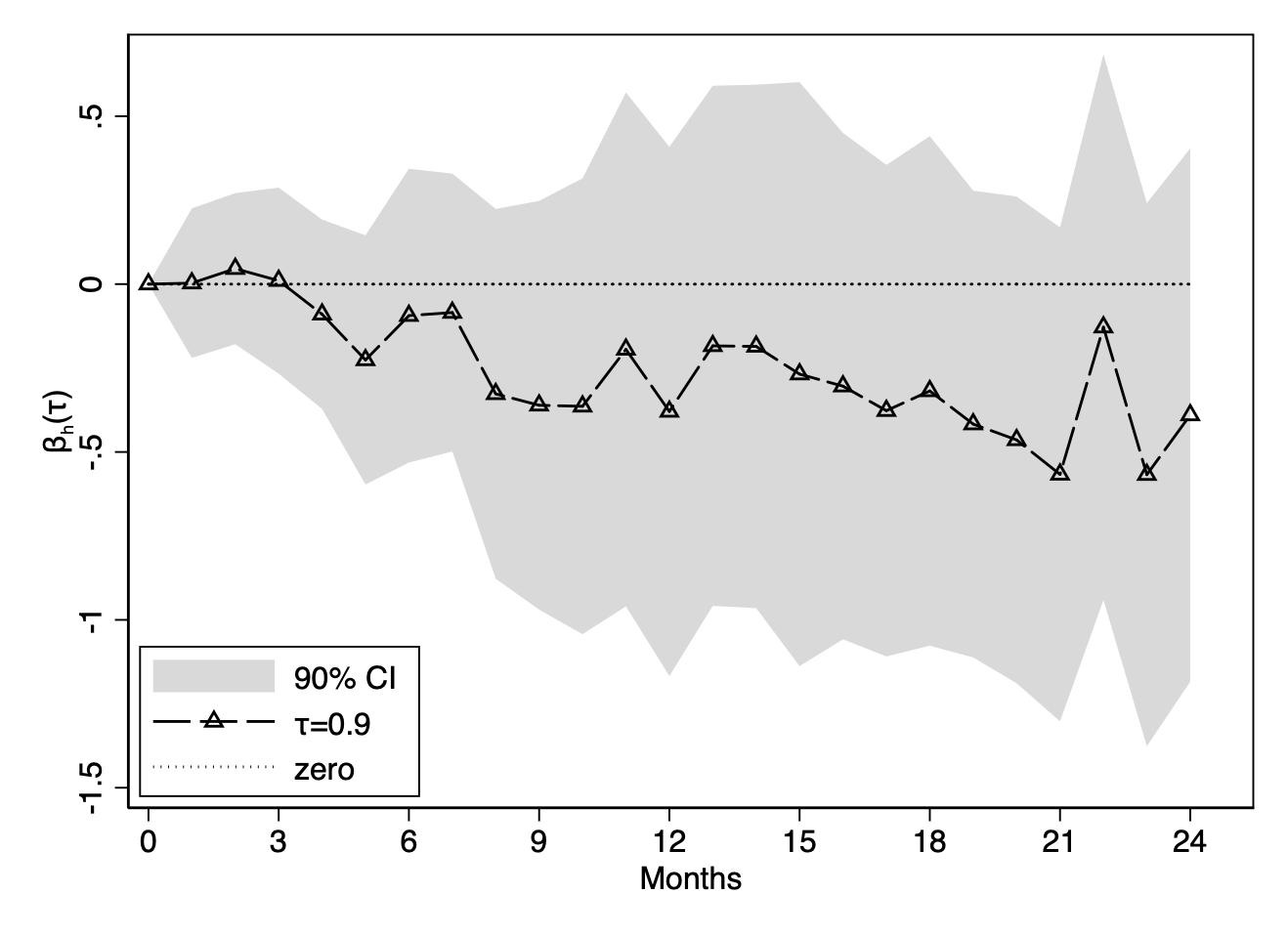}
	    \caption{ $\tau = 0.9$ with 90\% C.I.}
    \end{subfigure}
       \caption{Cumulative response of Industrial Production (in \% pts.) from a shock that increases volatility risk by one standard deviation, plotted for three quantiles $\tau \in \{0.1,0.5,0.9\}$. Y-axis is $\hat{\beta}_h(\tau)$, x-axis is the horizon $h$ in months. Dashed lines plot the quantile impulse response. Shaded area is the block-of-block bootstrap 90\% Confidence Interval (with block length of 7, and 1,000 bootstrap replications). Note that the impact response (horizon $h=0$) is by assumption zero, given our timing restrictions.}
\label{QIR_Vol}
\end{figure}

Figure \ref{QIR_Vol} shows the results of estimating the same model but using volatility risk in place of credit risk as the treatment variable. Comparing figure \ref{QIR_Vol} to figure \ref{QIR_ebp} suggests that the relationship between shocks to volatility risk and growth is similar to the relationship between shock to credit risk and growth. The timing and magnitude of the quantile impulse responses are almost identical following increases in volatility risk and credit risk. Both volatility and credit risk affect down-side more than upside-risk. The similarities are striking considering the fact that the sample correlation coefficient between these two variables is very low at $0.1$. We believe this is because overall financial conditions -- of which credit and volatility are both components -- have an asymmetric effect on the distribution of output growth.


\section{Conclusion}\label{sec:end}

Conventional econometric methods that model the mean impulse responses of growth to financial shocks can underestimate the true importance of financial shocks as causes of recessions. By now, this is widely appreciated by academics and policy-makers alike, which explains why a lot of research effort is put towards understanding the downside risks to growth. 

We offer a new methodology to identify the causal drivers of Growth-at-Risk. Our identification strategy is based on controls, yet it identifies  treatment effects on unconditional quantiles. In our view the distinction between conditional and unconditional quantiles of growth is important in the context of GaR. Conditionally low growth rates map to periods when the economy under-performs expectations, for example in a favorable macroeconomic climate this would mean high-yet-disappointing growth. On the other hand, unconditionally low growth rates always map to downturns and recessions, and as such are of primary concern for policymakers and academics. Our framework allows to study the latter while using familiar controls-based idenitification strategies based on timing restrictions.

Understanding the structural drivers of growth vulnerability can help discipline theoretical work and macroprudential policy efforts. Our empirical findings show that financial shocks have very large effects on downside risks with little upside. This suggests that stabilizing them can help avoid painful recessions, without large growth losses during the expansions.

\pagebreak
\printbibliography
\pagebreak

\pagebreak

\pagebreak

\begin{figure}[h!]
    \centering
    \begin{subfigure}[b]{0.49\textwidth}
        \centering
        \includegraphics[width=\textwidth]{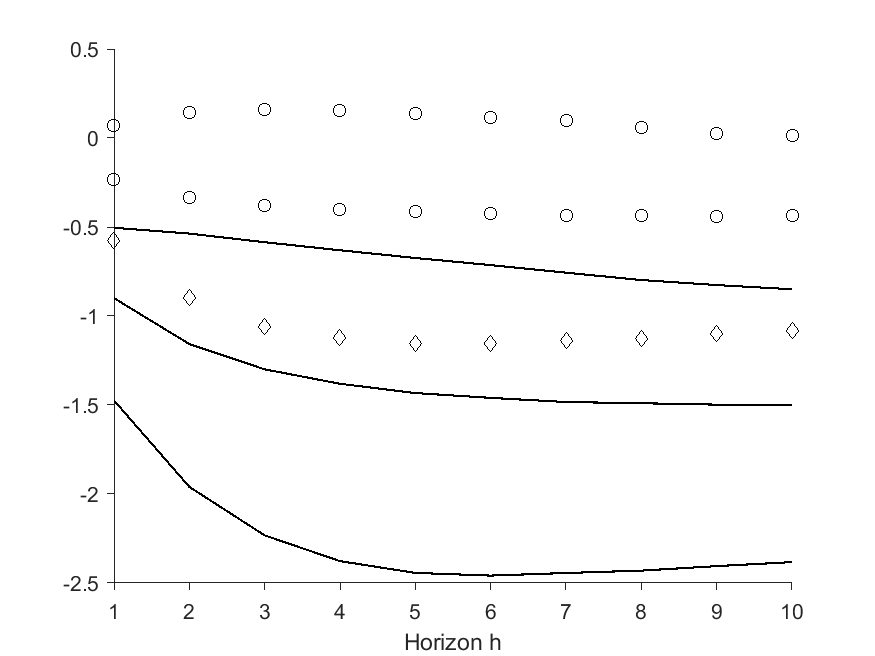}
        \caption*{QLP no controls}
    \end{subfigure}
    \begin{subfigure}[b]{0.49\textwidth}
        \centering
        \includegraphics[width=\textwidth]{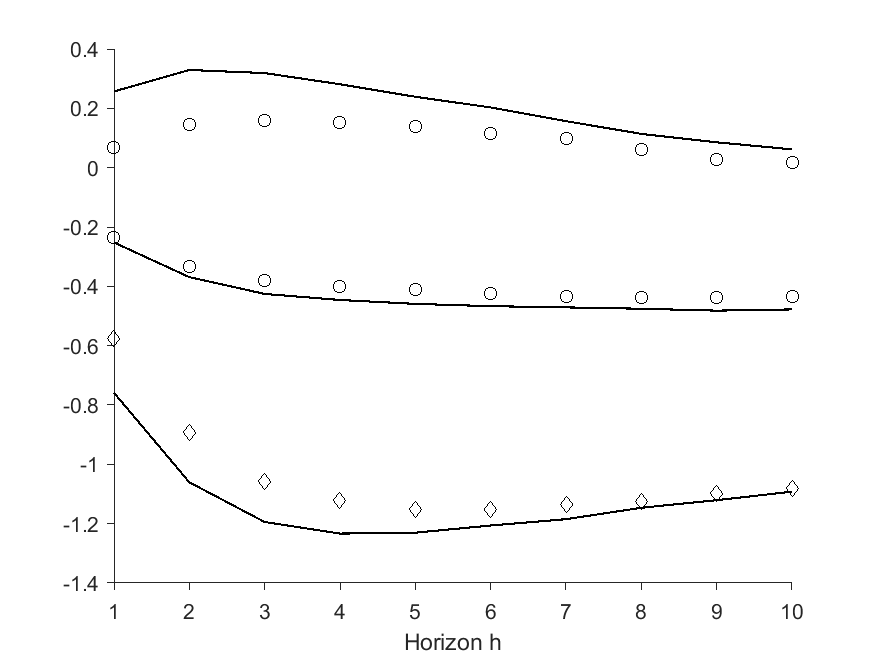}
        \caption*{QLP with controls}
    \end{subfigure}

    \begin{subfigure}[b]{0.49\textwidth}
        \centering
        \includegraphics[width=\textwidth]{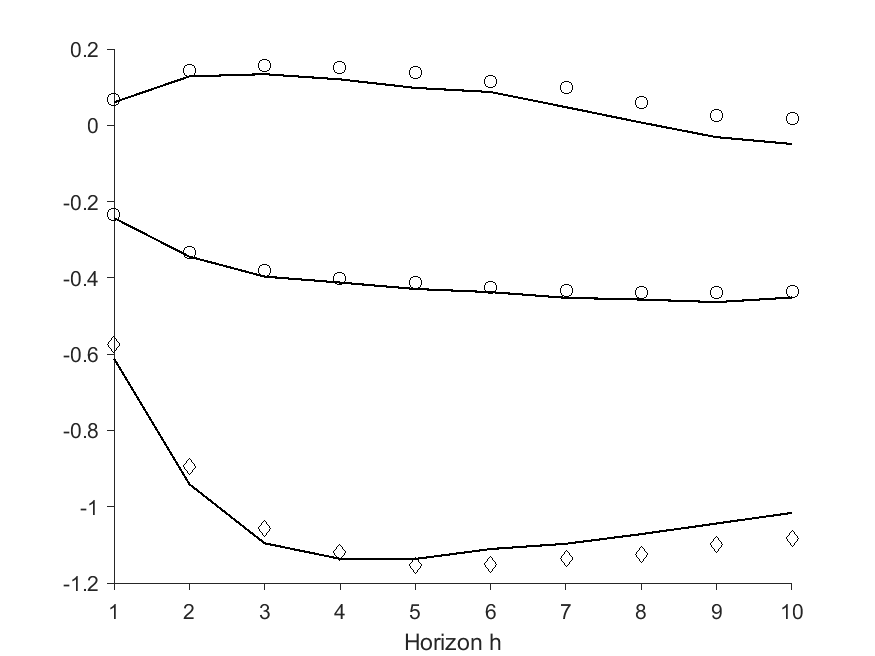}
        \caption*{GQR LP}
    \end{subfigure}
	\caption{Simulation results for the cumulative quantile impulse response of our illustrative example. The diamonds $ \diamond$, circles $\circ$ and triangles $\scalebox{0.6}{$\triangle$}$ are the same across the three panels and show the (linear approximation to the) true quantile impulse response as estimated by $Y_{t+h} = \alpha_h(U_{t+h}) +  \beta_h(U_{t+h}) Z^D_t $, for quantiles $\tau \in \{ 0.1 \diamond, 0.5 \circ ,0.9 \scalebox{0.6}{$\triangle$}  \}$. Solid lines show the results from the three estimators considered. QLP refers to the quantile local projection framework which uses the \cite{Koenker1978} estimator. GQR LP is our local projections based framework which uses the generalized quantile regression estimator introduced  by \cite{Powell2020}. Results are averaged over MC = 1,000 simulation replications, with a time-series of length T = 1,000 (after dropping 1,000 initial observations). Y-axis plots $\hat\beta_h(\tau)$ and x-axis shows the horizon $h$. }
    \captionsetup{labelformat=empty}
    \label{fig:MC}
\end{figure}

\pagebreak

\pagebreak

\pagebreak

\section{Appendix}

\subsection{Proofs}

We restate the theorem \ref{th:powell1} from \cite{Powell2020} in our settng. First, we want to show: $ \mathbb{P}[Y_{t+h} \leq q_h(\tau \mid D_t) \mid  D_t,  W^\top_t] = \mathbb{P}[Y_{t+h} \leq q_h(\tau \mid D_t)  \mid  W^\top_t]$. Evaluating the left hand side of the equality we have:
\begin{align*}
\mathbb{P}[Y_{t+h} \leq q_h(\tau \mid D_t) \mid  D_t,  W^\top_t] &= \mathbb{P}[Y_{t,h}(D_t) \leq q_h(\tau \mid D_t) \mid  D_t,  W^\top_t] \\ &= \mathbb{P}[Y_{t,h}(d) \leq q_h(\tau \mid d) \mid  D_t,  W^\top_t]  \\ &= \mathbb{P}[Y_{t,h}(d) \leq q_h(\tau \mid d) \mid  W^\top_t].
\end{align*}
The first equality sign follows from the definition of a potential outcome in assumption \ref{as:1}. The second equality sign comes from the rank similarity assumption \ref{as:3} which must hold for all $d, d'$ and thus also for $d = D_t$. The third equality sign follows from the conditional independence assumption \ref{as:3}. Evaluating the right hand side of the equality we have:
\begin{align*}
\mathbb{P}[Y_{t+h} \leq q_h(\tau \mid D_t) \mid  W^\top_t] &= \mathbb{P}[Y_{t,h}(D_t) \leq q_h(\tau \mid D_t) \mid  W^\top_t] \\ &= \int \mathbb{P}[Y_{t,h}(D_t) \leq q_h(\tau \mid D_t) \mid  W^\top_t, D_t] d\mathbb{P}(D_t \mid W^\top_t) \\  &= \int \mathbb{P}[Y_{t,h}(d) \leq q_h(\tau \mid d) \mid  D_t,  W^\top_t] d\mathbb{P}(D_t \mid W^\top_t) \\ &= \mathbb{P}[Y_{t,h}(d) \leq q_h(\tau \mid d) \mid  W^\top_t].
\end{align*}
The first equality follows from the definition of a potential outcome in assumption \ref{as:1}. The third equality follows from the rank similarity assumption \ref{as:4}. The second and fourth equality follow directly from properties of marginal probability functions.

Now we want to show: $\mathbb{P}[Y_{t+h} \leq q_h(\tau \mid D_t)] = \tau$.
\begin{align*}
\mathbb{P}[Y_{t+h} \leq q_h(\tau \mid D_t)] &= \int \mathbb{P}[Y_{t,h}(D_t) \leq q_h(\tau \mid D_t) \mid  W^\top_t, D_t] d\mathbb{P}(W^\top_t, D_t) \\  &= \int \mathbb{P}[Y_{t,h}(d) \leq q_h(\tau \mid d) \mid  W^\top_t, D_t] d\mathbb{P}(W^\top_t, D_t) \\  &= \mathbb{P}[Y_{t,h}(d) \leq q_h(\tau \mid d)] \\  &= \tau 
\end{align*}
The second equality follows from the rank similarity assumption \ref{as:3}. The fourth equality follows from  assumption \ref{as:1}. 

\pagebreak

\subsection{Illustrative Example simulation results}

\begin{table}[h!]
\begin{tabular}{cl|ll|ll|ll|lll|}
\multicolumn{1}{l}{}         &         & \multicolumn{2}{c}{QR no controls} & \multicolumn{2}{c}{QR with controls} & \multicolumn{2}{c}{GQR} \\
\multicolumn{1}{l}{Quantile} & Horizon & Mean Bias          & RMSE          & Mean Bias           & RMSE           & Mean Bias    & RMSE     \\
\hline
\multirow{10}{*}{0.1}        & 1       & -0.899             & 0.899         & -0.183              & 0.183          & -0.036       & 0.109    \\
                             & 2       & -1.066             & 1.066         & -0.166              & 0.179          & -0.047       & 0.177    \\
                             & 3       & -1.176             & 1.176         & -0.138              & 0.203          & -0.039       & 0.227    \\
                             & 4       & -1.256             & 1.257         & -0.113              & 0.239          & -0.016       & 0.253    \\
                             & 5       & -1.291             & 1.291         & -0.078              & 0.272          & 0.016        & 0.276    \\
                             & 6       & -1.307             & 1.308         & -0.054              & 0.308          & 0.041        & 0.302    \\
                             & 7       & -1.307             & 1.311         & -0.048              & 0.352          & 0.040        & 0.322    \\
                             & 8       & -1.306             & 1.316         & -0.022              & 0.389          & 0.052        & 0.351    \\
                             & 9       & -1.307             & 1.319         & -0.022              & 0.430          & 0.055        & 0.371    \\
                             & 10      & -1.300             & 1.316         & -0.009              & 0.455          & 0.067        & 0.386    \\
\hline

\multirow{10}{*}{0.5}        & 1       & -0.665             & 0.665         & -0.017              & 0.035          & -0.008       & 0.057    \\
                             & 2       & -0.823             & 0.823         & -0.034              & 0.075          & -0.010       & 0.093    \\
                             & 3       & -0.918             & 0.918         & -0.044              & 0.109          & -0.016       & 0.122    \\
                             & 4       & -0.980             & 0.980         & -0.045              & 0.140          & -0.012       & 0.149    \\
                             & 5       & -1.022             & 1.022         & -0.048              & 0.168          & -0.018       & 0.171    \\
                             & 6       & -1.035             & 1.035         & -0.042              & 0.188          & -0.013       & 0.195    \\
                             & 7       & -1.049             & 1.049         & -0.037              & 0.207          & -0.018       & 0.212    \\
                             & 8       & -1.052             & 1.052         & -0.038              & 0.226          & -0.019       & 0.231    \\
                             & 9       & -1.059             & 1.059         & -0.044              & 0.238          & -0.025       & 0.251    \\
                             & 10      & -1.064             & 1.064         & -0.042              & 0.253          & -0.016       & 0.261    \\
\hline

\multirow{10}{*}{0.9}        & 1       & -0.573             & 0.573         & 0.188               & 0.188          & -0.009       & 0.075    \\
                             & 2       & -0.682             & 0.682         & 0.185               & 0.188          & -0.016       & 0.109    \\
                             & 3       & -0.743             & 0.743         & 0.161               & 0.185          & -0.024       & 0.140    \\
                             & 4       & -0.783             & 0.783         & 0.129               & 0.190          & -0.032       & 0.165    \\
                             & 5       & -0.813             & 0.813         & 0.100               & 0.207          & -0.041       & 0.193    \\
                             & 6       & -0.828             & 0.829         & 0.088               & 0.226          & -0.027       & 0.210    \\
                             & 7       & -0.855             & 0.855         & 0.058               & 0.241          & -0.051       & 0.234    \\
                             & 8       & -0.858             & 0.860         & 0.054               & 0.265          & -0.053       & 0.263    \\
                             & 9       & -0.852             & 0.855         & 0.059               & 0.296          & -0.058       & 0.280    \\
                             & 10      & -0.866             & 0.870         & 0.045               & 0.314          & -0.066       & 0.294   
\end{tabular}
\caption{Simulation results for the cumulative QIR of the illustrative example (complementing figure \ref{fig:MC} in the main text). The ``true'' QIR to which the estimators were compared with is in fact a linear approximation obtained from the quantile local projection model $Y_{t+h} = \alpha_h(U_{t+h}) +  \beta_h(U_{t+h}) Z^D_t $. The true QIR was obtained by averaging the estimated $\beta_h(\tau)$ over the monte carlo replications. RMSE is the root mean squared error. QLP refers to the quantile local projection framework which uses the \cite{Koenker1978} estimator, GQR is the generalized quantile regression estimator introduced  by \cite{Powell2020}. Results are from MC = 1,000 simulation replications, with a time-series of length T = 1,000 (after dropping 1,000 initial observations).}
\end{table}

\pagebreak

\subsection{Additional figures}

\begin{figure}[H]
    \centering
    \begin{subfigure}[b]{0.49\textwidth}
        \includegraphics[width=\textwidth]{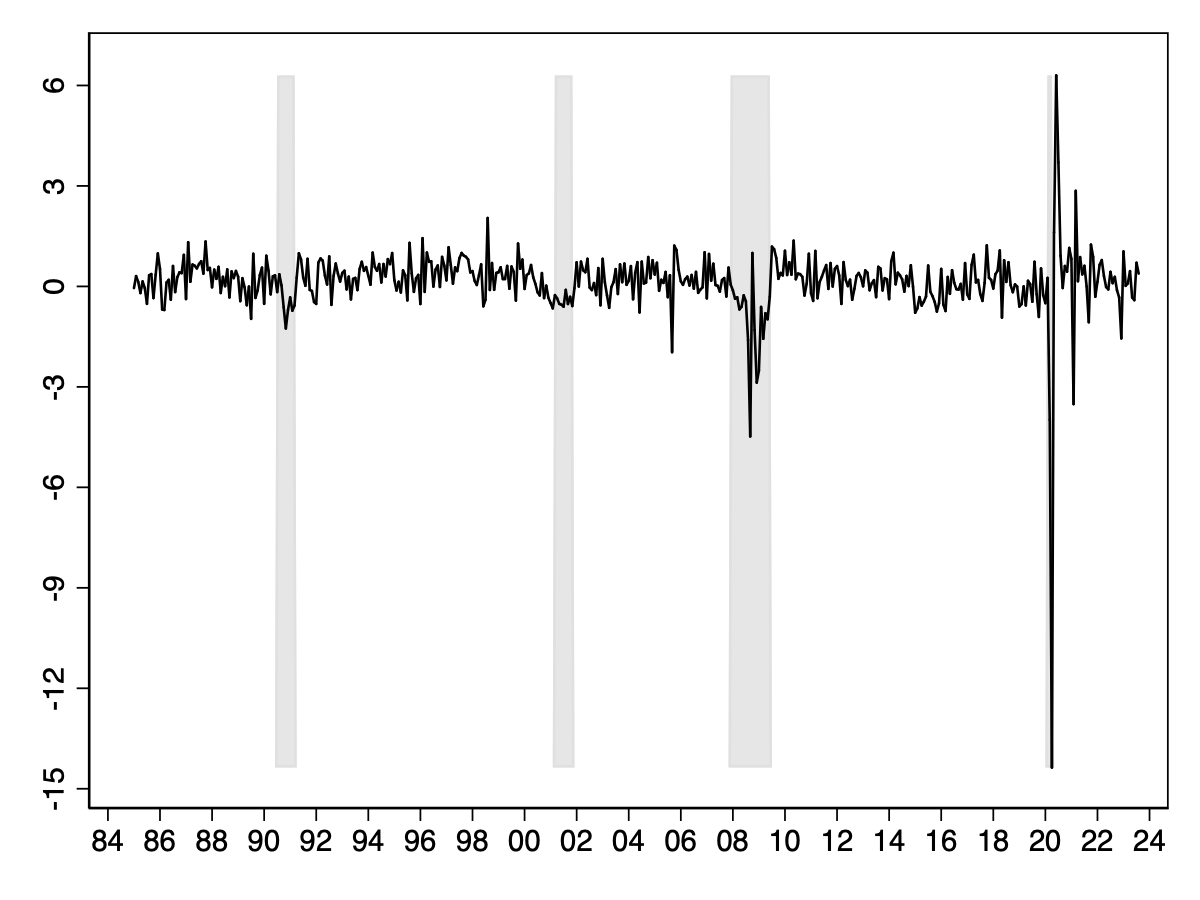}
        \caption{$Y_{t+1}$ - Monthly IP growth}
    \end{subfigure}
    \begin{subfigure}[b]{0.49\textwidth}
        \includegraphics[width=\textwidth]{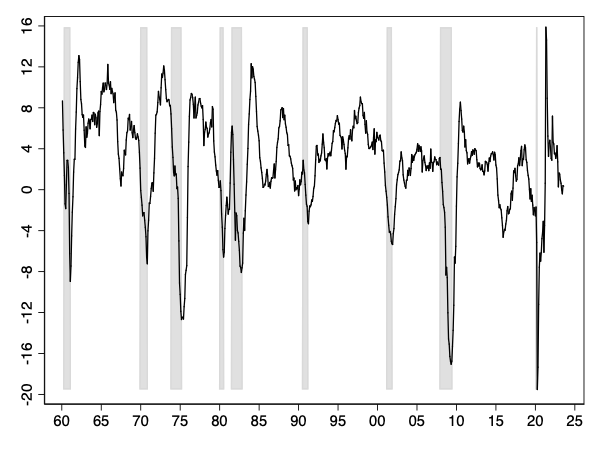}
        \caption{$Y_{t+12}$ - Annual IP growth}
    \end{subfigure}
\\
    \centering
    \begin{subfigure}[b]{0.49\textwidth}
        \includegraphics[width=\textwidth]{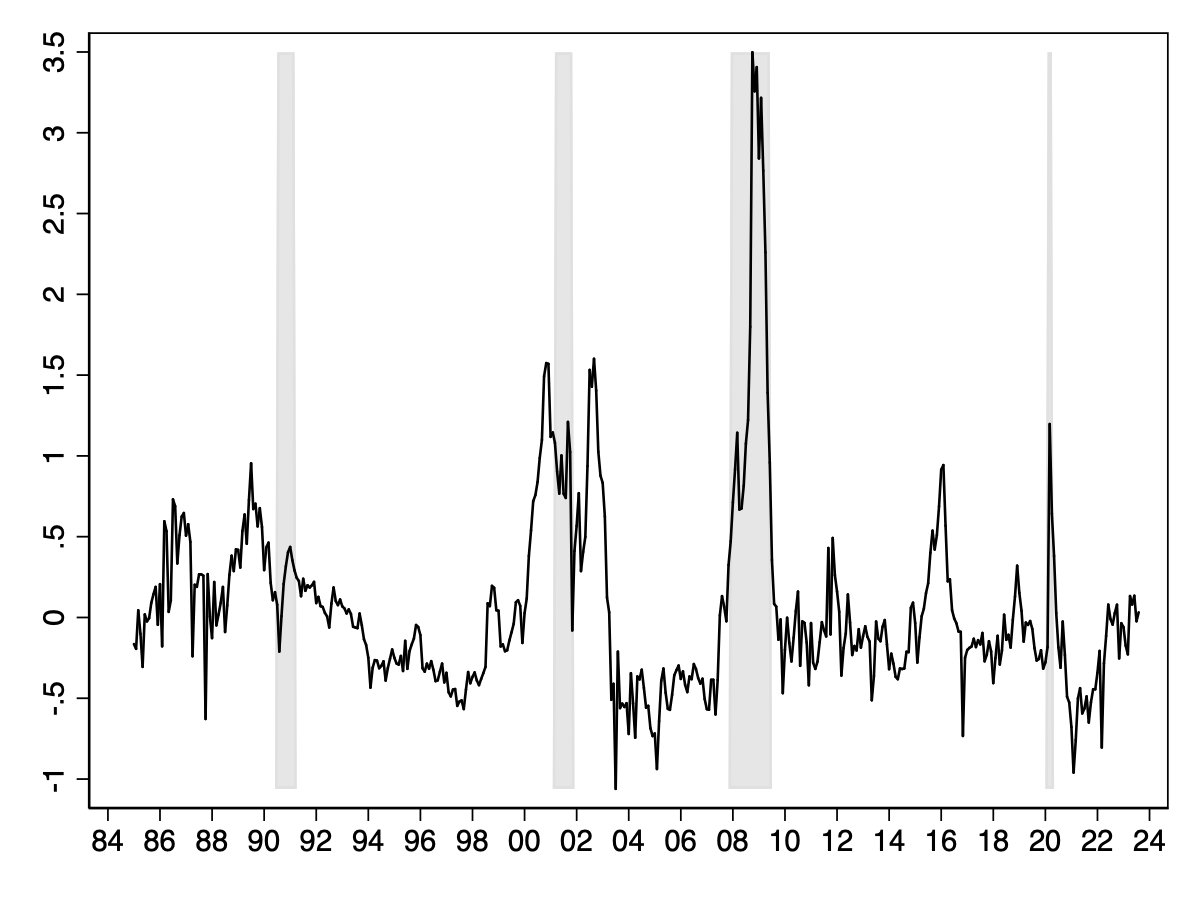}
        \caption{Excess Bond Premium}
    \end{subfigure}
    \begin{subfigure}[b]{0.49\textwidth}
        \includegraphics[width=\textwidth]{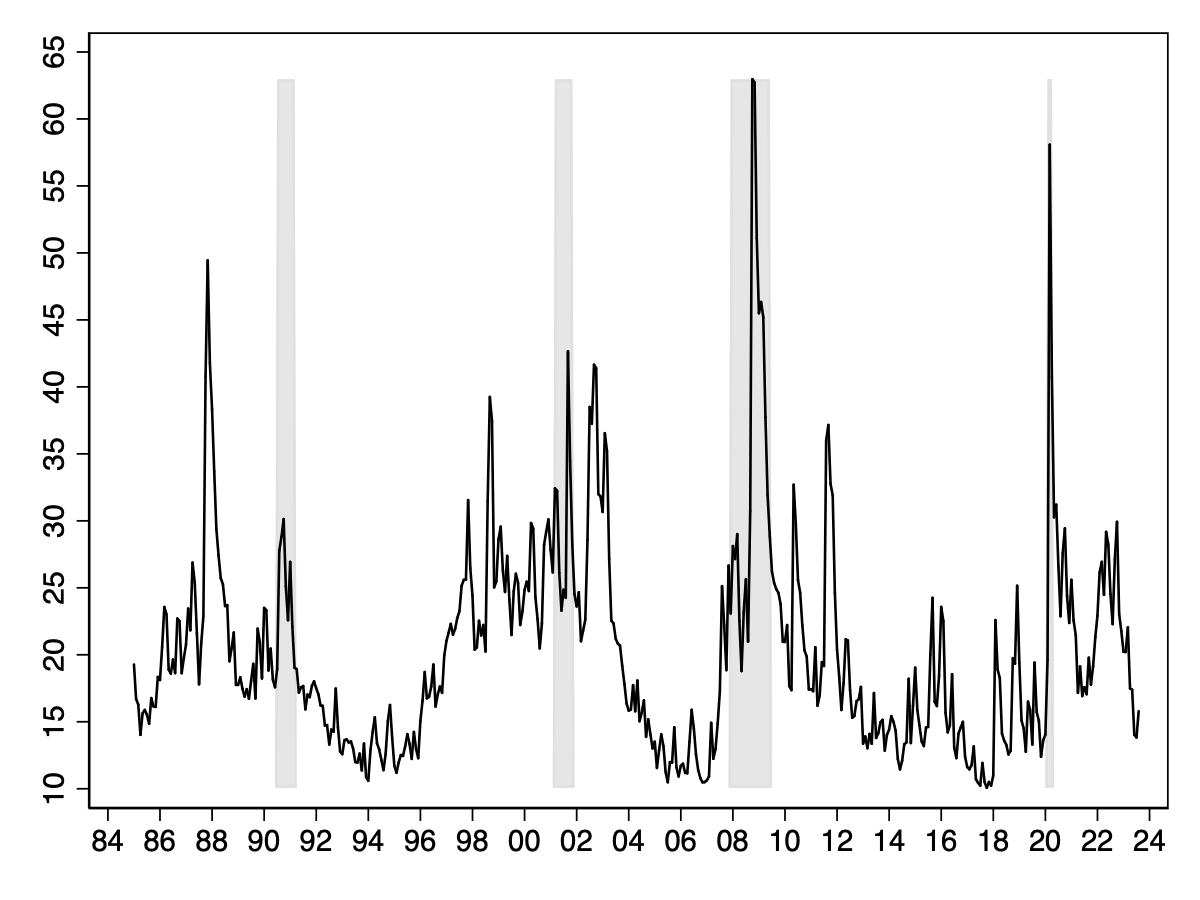}
        \caption{VIX Index}
    \end{subfigure}
\\
    \centering
    \begin{subfigure}[b]{0.49\textwidth}
        \includegraphics[width=\textwidth]{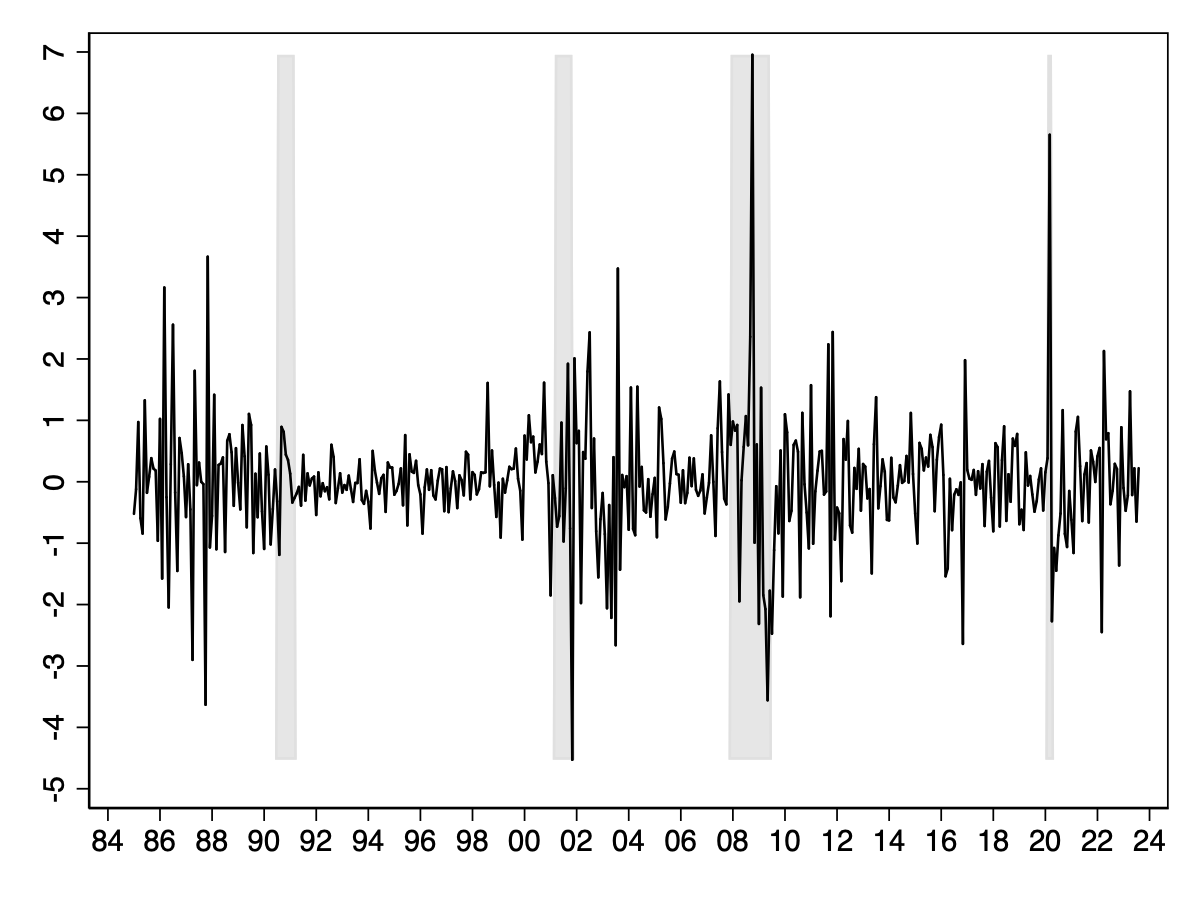}
        \caption{$D_t$ - Credit Risk}
    \end{subfigure}
    \begin{subfigure}[b]{0.49\textwidth}
        \includegraphics[width=\textwidth]{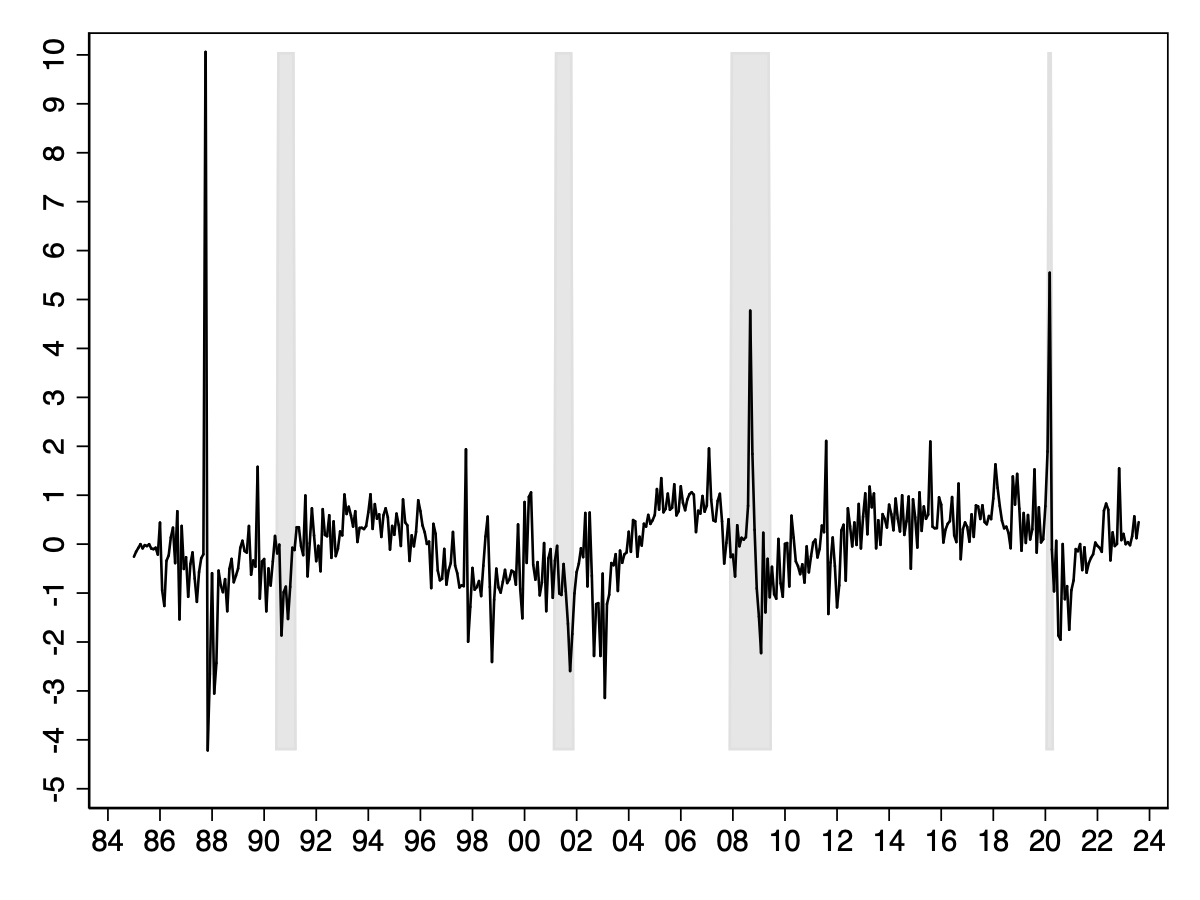}
        \caption{$D_t$ - Volatility Risk}
    \end{subfigure}
    \caption{Monthly time-series from January 1985 to August 2023. Grey bands indicate NBER recession dates. The series in the bottom panels have been Z-score normalized.}
    \label{fig:fourby}
\end{figure}

\pagebreak

\begin{figure}[H]
        \centering
        \begin{subfigure}[b]{0.49\textwidth}
            \centering
            \includegraphics[width=\textwidth]{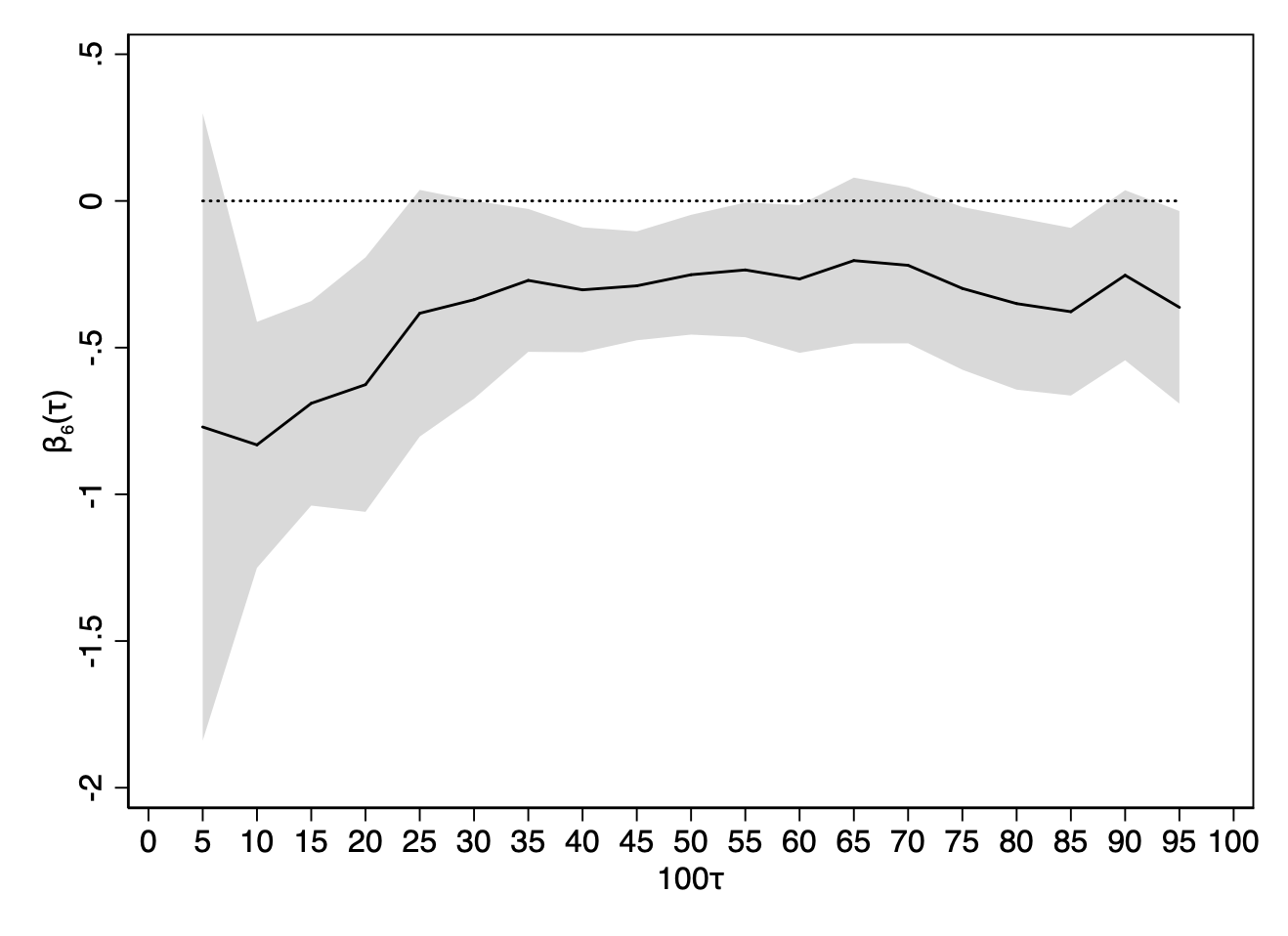}
	\caption{6 months}
        \end{subfigure}
        \hfill
        \begin{subfigure}[b]{0.49\textwidth}  
            \centering 
            \includegraphics[width=\textwidth]{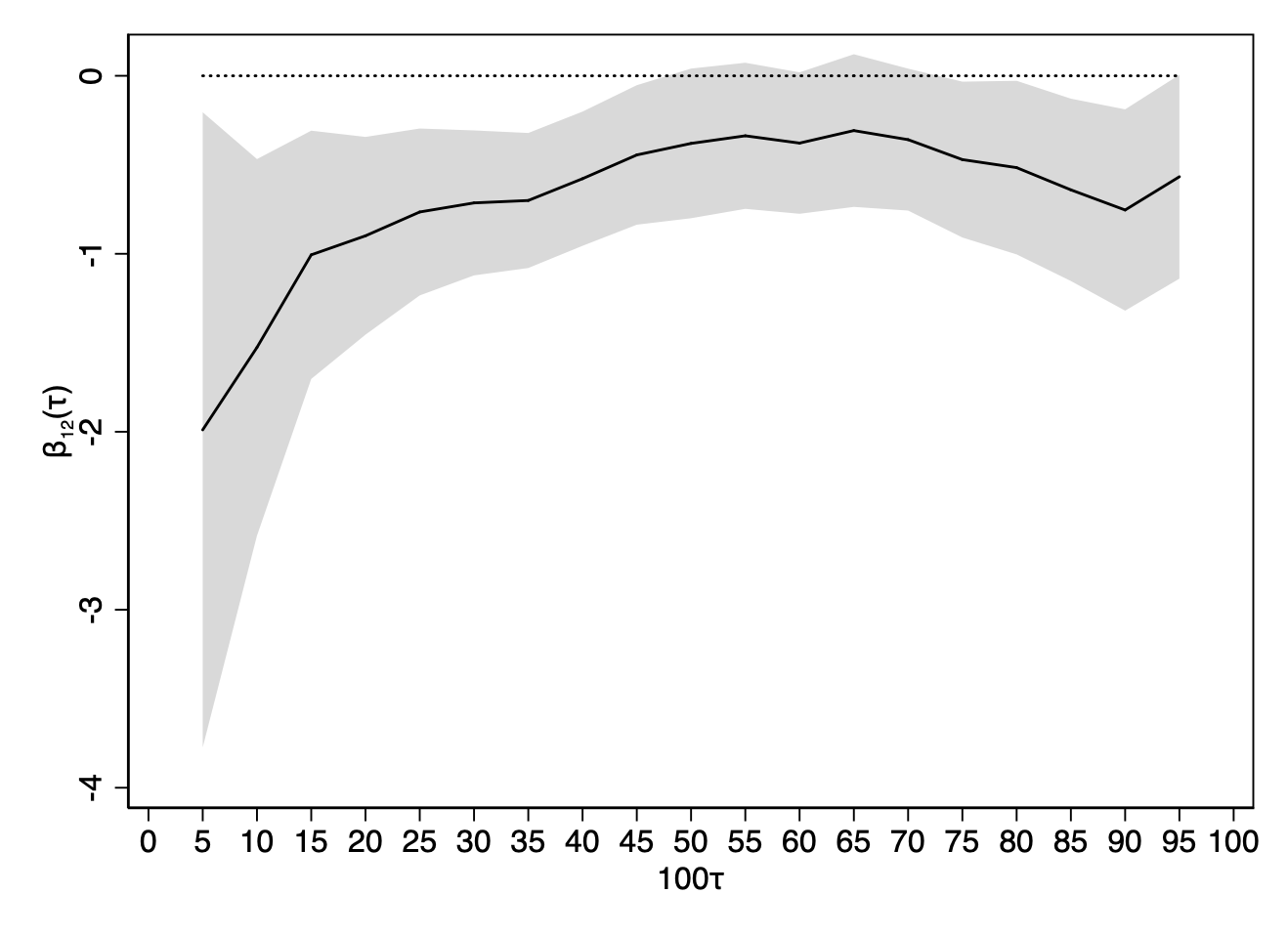}
\caption{1 year}
        \end{subfigure}

        \begin{subfigure}[b]{0.49\textwidth}   
            \centering 
            \includegraphics[width=\textwidth]{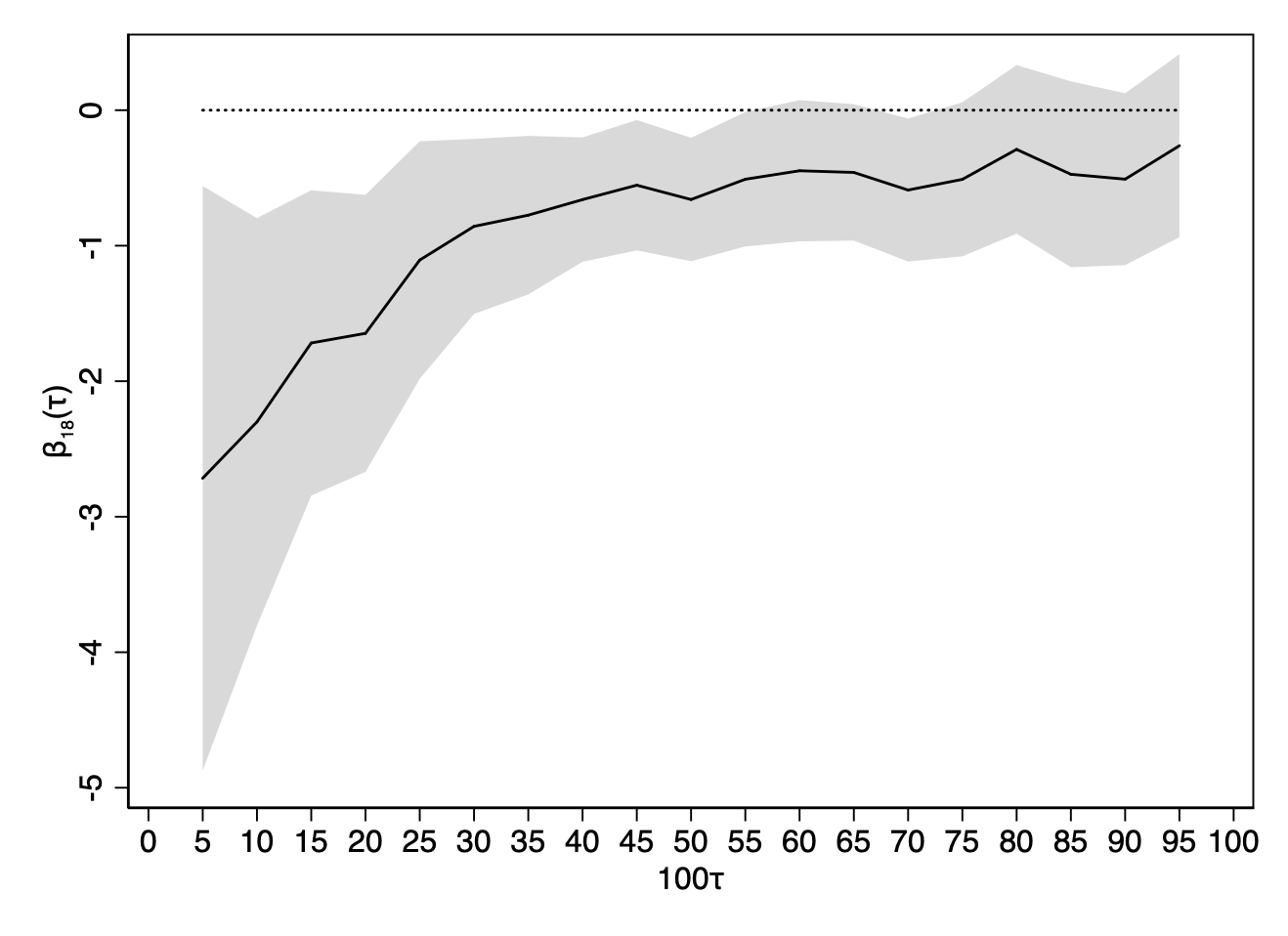}
\caption{18 months}
        \end{subfigure}
        \hfill
        \begin{subfigure}[b]{0.49\textwidth}   
            \centering 
            \includegraphics[width=\textwidth]{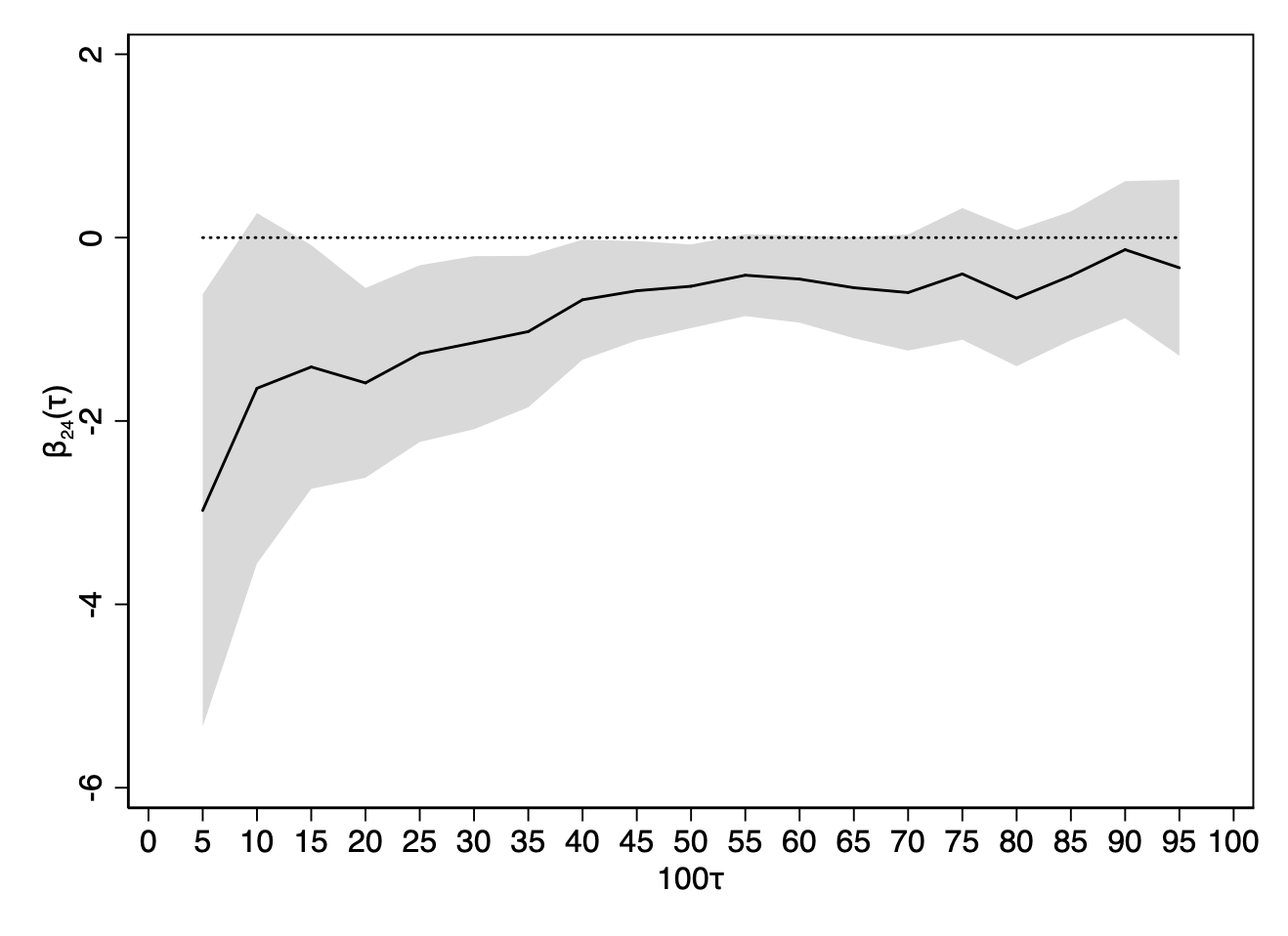}
\caption{2 years}
        \end{subfigure}
        \caption{Responses of Industrial Production (in \% pts.) to a shock that increases credit risk by one standard deviation, plotted for three horizons $h\in \{6,12,18,24\}$ (panels from top left to bottom right). The responses were estimated for quantiles from $\tau = 0.05$ to $\tau = 0.95$ in $0.05$ increments. Y-axis is the estimated response $\hat{\beta}_h(\tau)$, x-axis is the quantile $\tau$ (multiplied by $100$ for legibility). Shaded area is the block-of-block bootstrap 90\% Confidence Interval (with block length of 7, and 1,000 bootstrap replications).}
\label{many_q_ebp}
    \end{figure}

\pagebreak

\begin{figure}[H]
        \centering
        \begin{subfigure}[b]{0.49\textwidth}
            \centering
            \includegraphics[width=\textwidth]{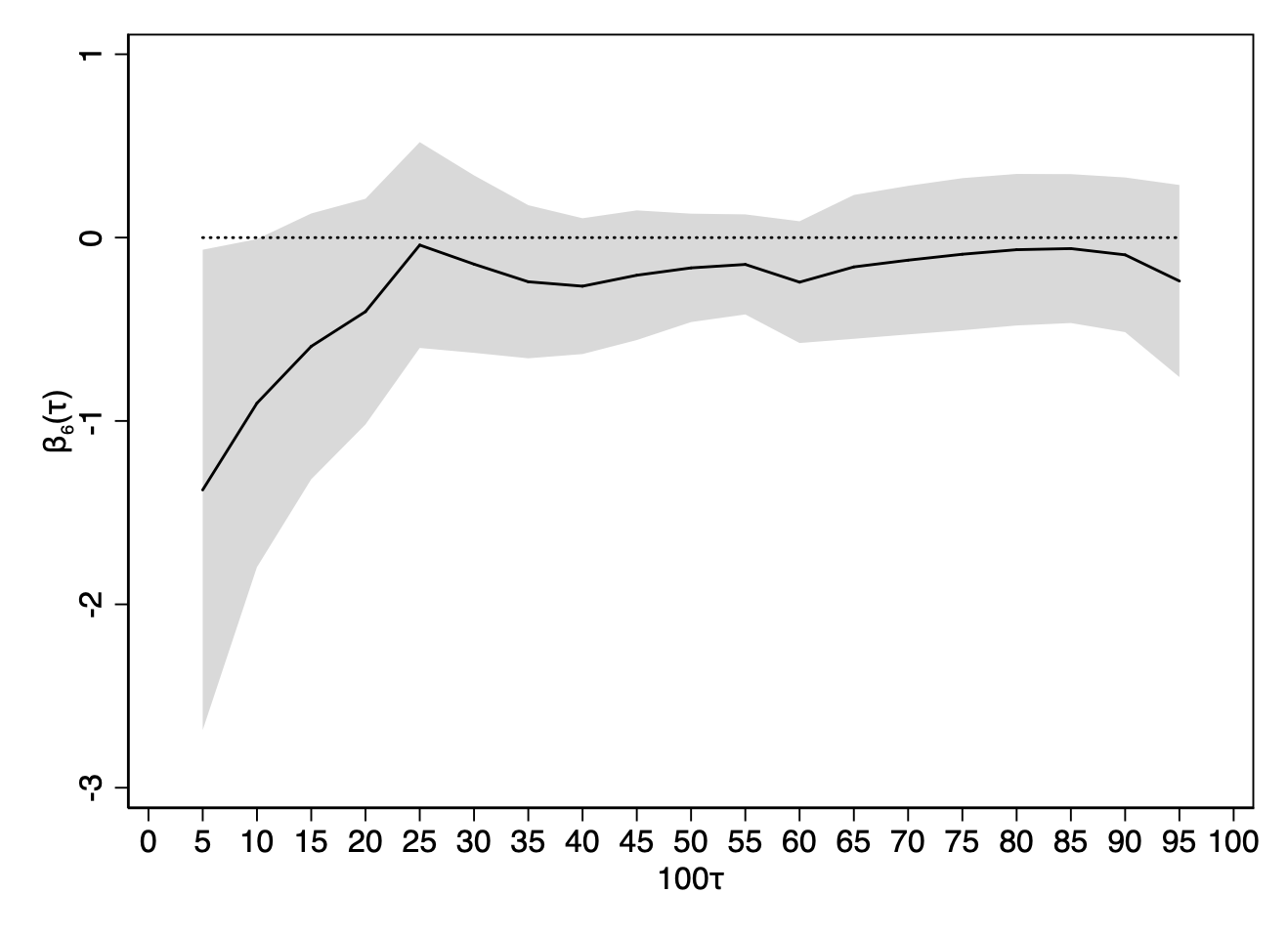}
	\caption{6 months}
        \end{subfigure}
        \hfill
        \begin{subfigure}[b]{0.49\textwidth}  
            \centering 
            \includegraphics[width=\textwidth]{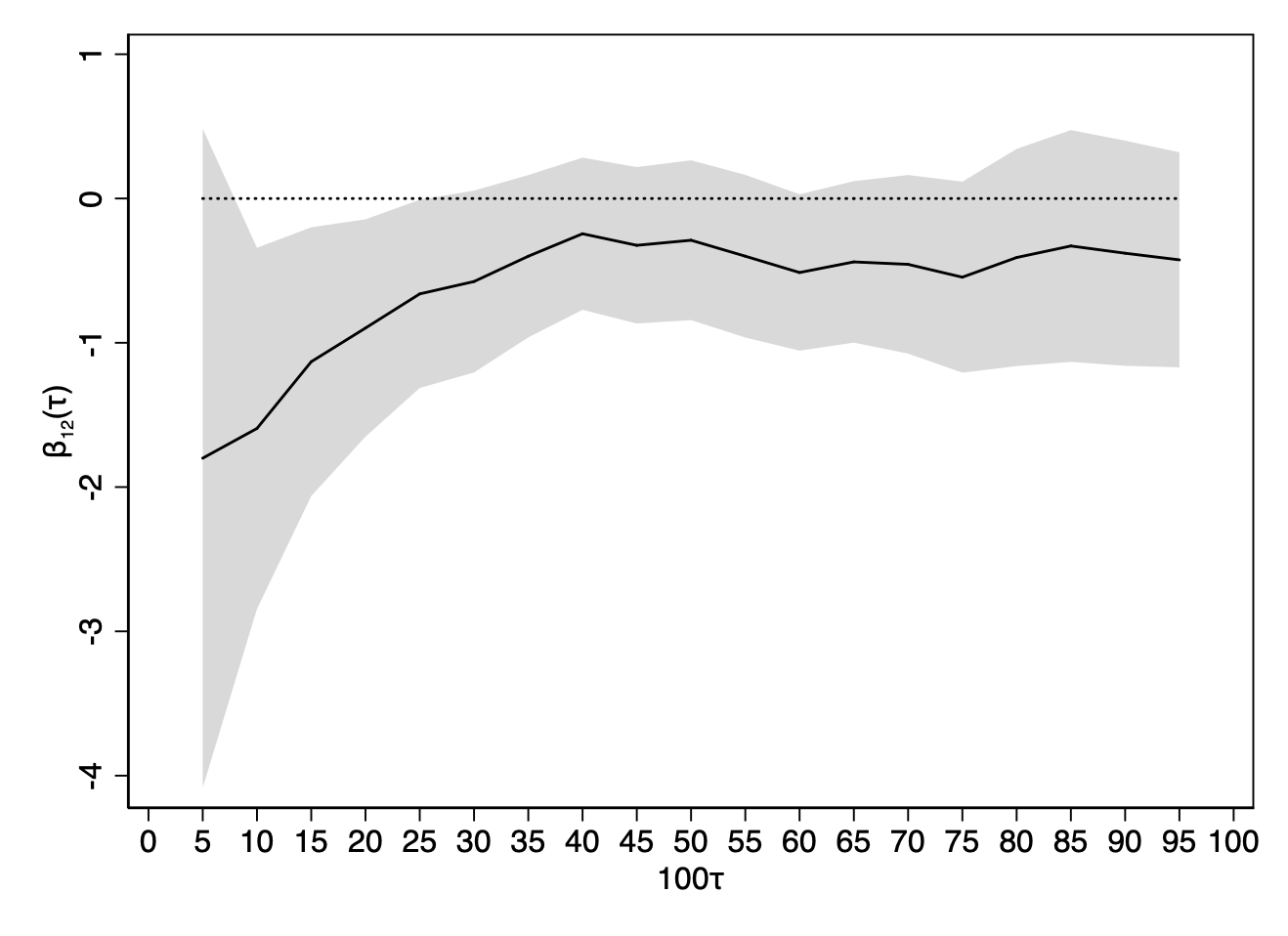}
	\caption{1 year}
        \end{subfigure}

        \begin{subfigure}[b]{0.49\textwidth}   
            \centering 
            \includegraphics[width=\textwidth]{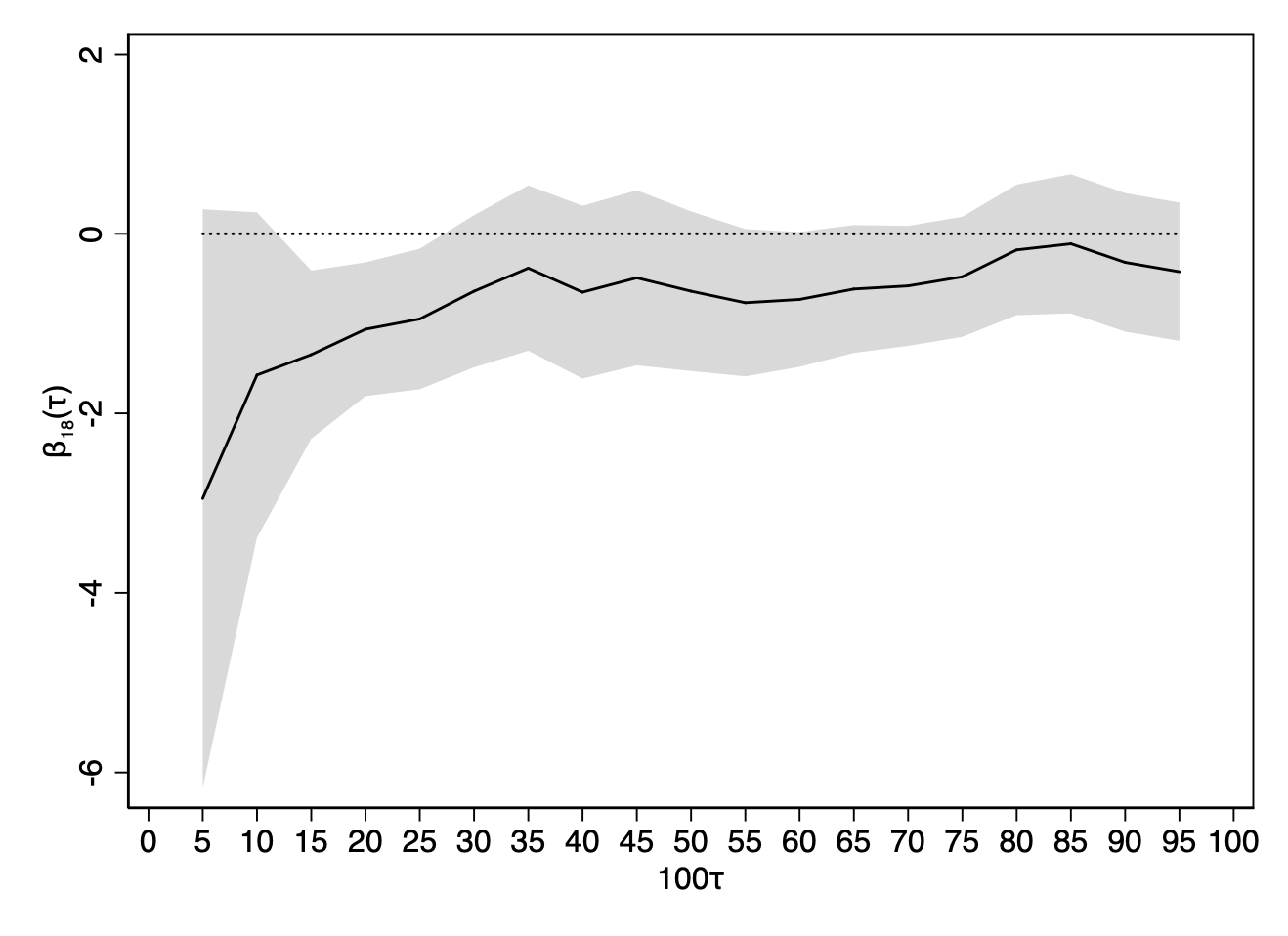}
	\caption{18 months}
        \end{subfigure}
        \hfill
        \begin{subfigure}[b]{0.49\textwidth}   
            \centering 
            \includegraphics[width=\textwidth]{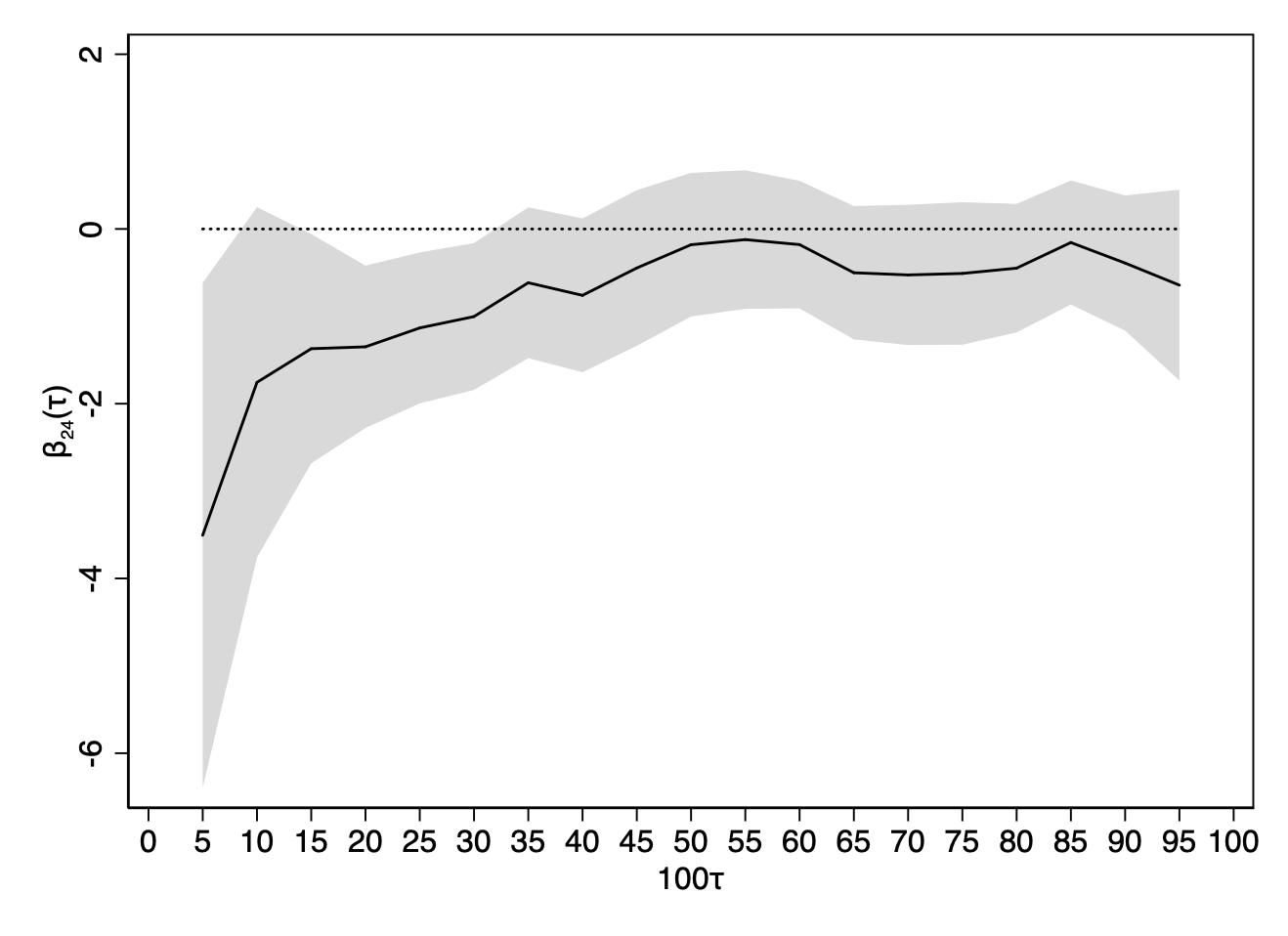}
	\caption{2 years}
        \end{subfigure}
        \caption{Responses of Industrial Production (in \% pts.) to a shock that increases volatility risk by one standard deviation, plotted for three horizons $h\in \{6,12,18,24\}$ (panels from top left to bottom right). The responses were estimated for quantiles from $\tau = 0.05$ to $\tau = 0.95$ in $0.05$ increments. Y-axis is the estimated response $\hat{\beta}_h(\tau)$, x-axis is the quantile $\tau$ (multiplied by $100$ for legibility). Shaded area is the block-of-block bootstrap 90\% Confidence Interval (with block length of 7, and 1,000 bootstrap replications).}
\label{many_q_vol}
    \end{figure}

\end{document}